\documentclass[aps,notitlepage,superscriptaddress]{revtex4}
\usepackage{multirow}
\usepackage{amsmath,amssymb,amsfonts,graphics,graphicx,dcolumn,bm}
\usepackage[ddmmyyyy,hhmmss]{datetime}

\begin{document}
\title{Socio-economic inequalities: a statistical physics perspective}

\author{Arnab Chatterjee}%
\email[Email: ]{arnabchat@gmail.com} 
\affiliation{Condensed Matter Physics Division, 
Saha Institute of Nuclear Physics, 1/AF Bidhannagar, Kolkata 700064 India.}


\begin{abstract}
Socio-economic inequalities are manifested in different aspects of our social life.
We discuss various aspects, beginning with the evolutionary and historical origins, 
and discussing the major issues from the social and economic point of view.
The subject has attracted scholars from across various disciplines, including physicists,
who bring in a unique perspective to the field. The major attempts to analyze the results, address the causes,
and understand the origins using statistical tools and statistical physics concepts are discussed.
\end{abstract}

\maketitle

\section{Introduction}
If you were in Dharmatola in Kolkata,  Manhattan in New York, Ikebukuro in Tokyo, Chandni Chowk in Delhi,
Normalm in Stockholm, Soho in London, Andheri in Mumbai, Kallio in Helsinki, Hutong in Beijing,
Montparnasse in Paris or San Salvario in Turin, 
you can observe different degrees of
inequality in socio-economic conditions, from standard of living, sanitation,
traffic, pollution, buying power or consumer behavior. 
The study of socio-economic inequalities has been important in identifying the causes 
to help minimize them by intervention of the state through laws and regulations.

Humans are social beings. Our social interactions are simple at times, but often more complex.
Social interactions in many forms produce spontaneous variations manifested as inequalities 
while at times these inequalities result out of continued complex interactions among the constituent human units.
With the availability of a large body of empirical data for a variety of measures from human social interactions,
it is becoming increasingly possible to uncover the patterns of socio-economic inequalities.
Additionally, the data can give us a clear picture of the origin of these variations 
as well as the dynamics of relevant quantities that produce these conditions, 
leading to their theoretical understanding.
This brief review aims at giving a unique perspective to the analysis
of complex social dynamics using the  statistical physics framework for
computational social 
science~\cite{lazer09}.
With tools of statistical physics as a core, the knowledge and 
techniques from various disciplines like statistics, applied mathematics,
information theory and computer science are incorporated for a better
understanding of the nature and origin of socio-economic inequalities that shape the humankind.

Sociology and Economics are disciplines in their own right, with a huge body of 
modern literature developed independently of physical sciences. 
However, in their infancy, these disciplines were not much distinct
from physical science. It is interesting to note that the development of statistical physics was also 
influenced by social statisticians recording the `state' of a person by recording the various
measures of his/her  social conditions. Of course, one of the most well known facts in the history of modern 
economics is that many of its ideas have been largely influenced by physical sciences, with their logical basis 
and technicalities having close resemblance with statistical physics.
A classic example is that of Jan Tinbergen, who with his colleague Ragner Frisch, 
was the first Nobel laureate in Economics (Nobel Memorial Prize in Economic Sciences  in 1969)
``for having developed and applied dynamic models for the analysis of economic processes''.
Tinbergen studied mathematics and physics at the University of Leiden under Paul Ehrenfest,
who was one of the key figures in the development of statistical physics. 
During his years at Leiden, Tinbergen had numerous discussions with 
Kamerlingh Onnes, Hendrik Lorentz, Pieter Zeeman, and Albert Einstein,
all Nobel laureates who left profound contributions to statistical physics.
Tinbergen and many other icons of modern economics shaped their ideas
with heavy influences from physical science, most of which were already  
developed in the literature of statistical physics.

From the other side, physicists seemed to have been thinking about 
socio-economic systems through the eyes of physical principles,
and had often been inspired from those problems to lay foundations 
to fields in mathematics and physics which became further relevant in
broader contexts.
Aristotle (384-322 BCE) is known to have laid the first foundations of physics
in his \textit{Physica}, where he even argued that barter is the form of exchange 
which is used in most societies.
Huygens (1629-1695), known for his contribution to optics, had laid the foundations of 
probability theory while working on insurance and annuities.
Astronomer Halley (1656-1742) was studying mortality and life expectancy besides studying celestial bodies.
Quetelet (1796-1874) was the among the first to study the average man (\textit{l'homme moyen}),
and describe different facets of human social life through statistics, laying the foundations
of what he termed as `social physics', while his contemporary Comte (1798-1857) provided solid foundations 
to the subject~\cite{ball04}, also calling it `sociology'.
Poincar\'{e}'s student Bachelier~\cite{bachelier1900theorie} studied the dynamics of stock market, was the first
to use Brownian motion to study stock price fluctuations, and predated Einstein's work on Brownian motion.
Other notable discussions include Majorana's article on the value of statistical laws in physics and in
social science~\cite{di1942valore}.
Kadanoff's work on urban dynamics~\cite{kadanoff1971simulation} was an interesting
study on the labor and management aspects of cities. 
In subsequent years, physicists became more interested in social phenomena, asking
interesting questions and proposing models to explain interesting aspects of social life
(See e.g., Montroll \& Badger~\cite{montroll1974introduction} as well as Ref.~\cite{anderson1988economy}),
which gave alternative views to the traditional approaches developed in mainstream social sciences.

Social scientists and economists have  dealt with numerous
interesting issues of the human society, uncovering behaviors which seem
to be universal. However, it is often pointed out that certain patterns 
observed a couple of decades ago may have changed. This happens due to 
the essential nature of human interaction which has changed with the
advent of technologies. This makes  socio-economic systems distinct 
compared to physical systems -- here we rarely find established `laws'~\cite{ball04}.
This naturally calls for a change in the theoretical analysis
and modeling, making the field an extremely challenging one.
Specifically, there has been  wide discussions about addressing
socio-economic phenomena in the light of 
\textit{complexity science}~\cite{buchanancomplexity}
and embracing ideas from various 
disciplines~\cite{buchanan07}
to understand socio-economic phenomena
and a few big initiatives are already taking shape~\cite{INET}.

Socio-economic inequality is the existence of unequal opportunities and rewards for various social
positions or statuses within the society. It usually contains structured and recurrent
patterns of unequal distributions of goods, wealth, opportunities, and
even rewards and punishments.
There are two main ways to measure social inequality: \textit{inequality of conditions},
and \textit{inequality of opportunities}.
\textit{Inequality of conditions} refers to the unequal distribution of income,
wealth and material goods. 
\textit{Inequality of opportunities} refers to the unequal distribution of `life
chances' across individuals. This is reflected in measures such as level of
education, health status, and treatment by the criminal justice system.
In this review, we will also discuss topics which are manifestations of unequal choices
made by groups of individuals, e.g., consensus, religion, bibliometric data etc.

With rapid technological developments and the world getting more and more connected,
our social conditions are experiencing changes, rapid in some parts
of the world, and much slower elsewhere. The nature of complexity of human
social interactions are also changing. The social classes formed out 
of societal evolution are changing, making it much difficult to predict
or even comprehend the near future.

In this review, we concentrate on a handful of issues, which have been of concern to
physicists in the same way as they have been drawing attention from social scientists.
There have been even deeper and sensitive issues in society, such as gender inequality,
racial or ethnic inequality, inequalities in health and education etc.~\cite{neckerman2004social} 
which we do not discuss here.
This review is organized as follows.
In Sec.~\ref{sec:evolution} we try to discuss the evolutionary origins of
socio-economic inequalities. We try to address the question whether
statistical physics can be successfully used to address the problems and model
situations that lead to such inequalities in Sec.~\ref{sec:statphys}.
In Sec.~\ref{sec:process} we will discuss a number of processes that are
relevant in our understanding of the phenomena. The most prominent issue
of income and wealth inequality will be discussed in Sec.~\ref{sec:income}.
Urban systems will be addressed in Sec.~\ref{sec:urban},
consensus in Sec.~\ref{sec:consensus}, bibliometrics in Sec.~\ref{sec:biblio}.
We will briefly discuss how networks are important in many contexts in Sec.~\ref{sec:networks}.
In Sec.~\ref{sec:measure} we present the different ways to measure inequalities, 
and Sec.~\ref{sec:deal} will discuss some possible ways to deal with socio-economic inequalities.
Finally we conclude with discussions in Sec.~\ref{sec:discuss}.

\section{Evolutionary perspective to socio-economic inequality}
\label{sec:evolution}
The human species has lived as hunter gathers for more than 90\% of its existence. 
It is widely known that hunter-gatherer societies were egalitarian, as is still evident from
the lifestyle of various tribes likes the !Kung people of the Kalahari desert~\cite{pennisi2014our}. 
They traditionally had few possessions, and were semi nomadic in the sense they were moving periodically.
They hardly mastered farming and lived as small groups. The mere instinct to survive
was driving them to suppress individual interests.
They shared what they had, so that all of their group members were healthy and strong,
be it food, weapons, property, or territory~\cite{buchanan2014more}.
When agricultural societies developed, it created elaborate hierarchies,
which certainly had more or less stable leadership in course of time. 
These evolved into clans or groups led by family lines, which eventually blew up as kingdoms.
In these relatively complex scenarios, chiefs or kings assumed new strategies for hoarding surplus
produce of agriculture or goods, predominantly for survival in times of need,
and thus concentrating wealth and power 
(see Sec.~\ref{sec:model_wealth} for models with savings, also Ref.~\cite{Chatterjee-2007}). 
There were mechanisms involved that
helped in wealth multiplication, along with the advancement of technologies.
The transition from egalitarianism to societies with competition and inequality
paved way for the development of chiefdoms, states and industrial empires~\cite{pringle2014ancient}.

\section{Why statistical physics?}
\label{sec:statphys}
Given that many laws of nature are statistical in origin,  a well established
fact that applies across most areas of modern physics, gives statistical physics
a status of a prominent and extremely useful discipline.
The subject, being developed in a very general framework,
makes it applicable to various areas outside the commonly perceived
boundaries of physics, as in biology, computer science and information technology,
geosciences etc. It comes as no big surprise that attempts to understand the nature of
social phenomena using statistical physics has been going on since some time, with the research activity gaining
much more prominence in the recent two
decades. What is much more striking is that
the usefulness of the statistical physics to understand social systems has gone
far from being merely `prominent' to being 
`important'~\cite{chakrabarti2013econophysics,Castellano:2009,stauffer2006biology,stauffer2013biased,galam2012sociophysics,Sen:2013}.
Statistical physics treats its basic entities as particles, while in society the basic
constituents are human individuals. In spite of this gross difference, human society shows remarkable
features which are similar to those widely observed in physical systems. 
Transitions are observed from order to disorder and emergence of consensus on a specific issue from a 
situation with multiple options by converging to one or a very few of them. 
Many measurables in a society show scaling and universality~\cite{buchanan07}.
Dynamical socio-economic systems exhibit 
self-similarity~\cite{sinha2010econophysics}
while going from one phase to another,
similar to critical phenomena observed in physical systems.
Statistical physics can be used as a tool to effectively understand the regularities
at a larger scale, that emerge out of interaction among individual units, the social being.

Statistical physics tells us that systems of many interacting dynamical units collectively exhibit, a behavior which is
determined by only a few basic dynamical features of the individual units and of the embedding dimension, 
but independent of all other details.
This  feature which is specific to critical phenomena, like in continuous phase transitions, is known as 
universality~\cite{stanley1971introduction}. 
There is enough empirical evidence that a number of social phenomena are characterized by simple emergent behavior out of the
interactions of many individuals. 
In recent  years, a growing community of researchers have been analyzing large-scale 
social dynamics to uncover universal patterns and also trying to propose simple microscopic models to describe them, 
similar to the minimalistic models used in statistical physics.
These studies have revealed interesting patterns and behaviors in social systems, e.g.,
in elections~\cite{fortunato2007scaling,chatterjee2013universality,mantovani2011scaling}, 
growth in population~\cite{rozenfeld2008laws} and economy~\cite{stanley1996scaling},
income and wealth distributions \cite{chakrabarti2013econophysics}, 
financial markets~\cite{mantegna2000introduction}, languages~\cite{petersen2012statistical}, etc.
(see Refs.~\cite{Castellano:2009,Sen:2013} for reviews).

\section{Processes leading to inequality and  broad distributions}
\label{sec:process}
Can we point out the key processes that make the probability distribution of a quantity broad?
Actually, there are several processes known to do so, and can be grossly characterized into a few
categories. In this review, we will discuss a few, that are most relevant to the issue
of socio-economic inequalities.

\subsubsection*{Random walk}
For a random walk of unitary steps, the probability of finding the walker at distance $x$ from origin after $N$ steps is given by
\begin{equation}
 P(x) = \frac{2}{\sqrt{\pi} N}  \exp\left(-\frac{x^2}{2N}\right).
 \label{eq:rw}
 \end{equation}
In case the steps are variable, with $l$ being the average length of the path, Eq.~(\ref{eq:rw})
gets modified to
\begin{equation}
 P(x) = \frac{2}{\sqrt{2\pi Nl^2}} \exp \left(-\frac{x^2}{2Nl^2}\right),
 \label{eq:rw1}
\end{equation}
still  a Gaussian distribution about the mean displacement.

Let us assume a random walk starting from origin. The time (number of steps) to return to origin
$t$ is even and we denote by $2N$, and let the probability of this event be $u_t$, and 
$f_t$ be the probability that the first return time is $t$.
One can easily write that the probability $u_{2N}$ that the walker is at position zero after $2N$
steps as
\begin{equation}
 u_{2N} =2^{-2N} \binom {2N}{N}
\end{equation}
from where it can be shown that the distribution of the first return times is
\begin{equation}
 f_{2N} = \frac{\binom{2N}{N}}{(2N-1)2^{2N}}.
\end{equation}
Now, $\ln f_{2N} = \ln (2N)! - 2\ln N! -2N \ln 2 - \ln(2N-1)$. Using Sterling's approximation
for $N \to \infty$ and simplifying, one can obtain
\begin{equation}
f_{2N} \simeq \sqrt{ \frac{2}{N(2N-1)^2}}  \sim N^{-3/2},
\end{equation}
or equivalently, $f_t \sim t^{-3/2}$, 
i.e., the distribution of return times has a power law decay with exponent $\frac{3}{2}$.

\subsubsection*{Combination of exponentials}
Let a certain quantity $y$ has an exponential distribution $P(y) \sim \exp(ay)$.
If the constant $a>0$, $P(y)$ should have an upper cutoff given by the maximum value of $y$ so that
the distribution is normalizable. Now let another quantity $x$ is related to $y$ as
$x \propto \exp(by)$, where $b$ is another constant. Then, one can write
\begin{equation}
 P(x) = P(y) \frac{dy}{dx} \sim \frac{\exp(ay)}{b\exp(by)} = \frac{x^{\frac{a}{b}}}{bx} = \frac{1}{b} x^{-(1-\frac{a}{b})}.
\end{equation}

\subsubsection*{Self-organized criticality}
In Self-Organized Criticality~\cite{bak1997nature}, the steady state of the evolution for 
a dissipative dynamical system is a fractal. The evolution of the piles of sand inspired this model, known as `sandpile' models. 
The sites can support grains of sand until a certain threshold, beyond which the site
becomes``critical'' and expels the sand to the neighbors. It may happen that
many sites can become critical and adding a single grain can create a large
scale avalanche. At criticality, the size of the avalanches are power law distributed,
and a feedback mechanism ensures that whenever the system is full of sand no more grain can be added
until the activity stops, while at times of no activity sand grains need to be added.
Thus the amount of sand fluctuates around a mean value, as a result of a dynamical equilibrium.

\subsubsection*{Multiplicative processes}
Let us assume that a quantity $x$  evolves in time with a multiplier w.r.t. its state in the previous step.
Formally, $x_{t+1} = \epsilon_t x_t$. Had it evolved from $t=0$, one can write
\begin{equation}
 x_{t+1} = \left[ \prod_{i=0,1,\ldots,t} \epsilon_i \right]x_0,
\end{equation}
where $\epsilon_t$ is a random multiplier changing with time $t$.
Taking logarithms, 
\begin{equation}
 \log x_{t+1} =  \log x_0 + \sum_{i=0,1,\ldots,t} \log \epsilon_i.
\end{equation}
The second term, which is the sum of random variables in a log scale will simply
produce a normal distribution independent of the precise nature of 
the distribution of the $\epsilon$. Hence, the variable $x$ follows 
the `lognormal' distribution
\begin{equation}
  P(x) 
  =
  \frac{1}{x \sqrt{2\pi \sigma^2}} 
  \exp{\left\{ -\frac{(\log x -\mu)^2}{2\sigma^2}\right\}} \, ,
  \label{eq:lognormal}
\end{equation}
where $\mu = \langle \log(x) \rangle $ is the mean value 
of the logarithmic variable 
and $\sigma^2 = \langle [\log(x) - \mu]^2 \rangle $ 
the corresponding variance. 
However, there is often a confusion in the range where the variance is quite large $\sigma^2 \gg (\log x -\mu)^2$,
so that  $P(x) \sim 1/x$ in the leading order term.

A small change in the generative mechanism can also produce a power law distribution, which was nicely
demonstrated by Champernowne~\cite{Champernowne:1953} in the context of the power law
tail of income distribution (which will be discussed in the following section).
Champernowne assumed that there is some minimum income $m$.
For the first range, one considers incomes between $m$ and $\gamma m$,
for some $\gamma > 1$;
for the second range, one considers incomes between $\gamma m$ and
$\gamma^2 m$. A person is assumed to be in class $j$ for $j \ge 1$
if their income is between $m \gamma^{j-1}$ and $m \gamma^j$.
It is  assumed that over each time step,
the probability of an individual moving from class $i$ to class $j$,
denoted by $p_{ij}$, depends only on the value of $|j-i|$.
The equilibrium distribution of individuals among classes under this
assumption gives Pareto law. 
It is important to note that in multiplicative models, 
income becomes arbitrarily close to zero due to successive multiplications,
while in case of the Champernowne model, there is a
minimum income corresponding to the lowest class which is a bounded minimum that
acts as a reflective barrier to the multiplicative model. This produces a power
law instead of a lognormal distribution~\cite{mitzenmacher2004brief}.

\subsubsection*{Preferential attachment}
In one of the most prominent works to discuss preferential attachment,
Simon~\cite{simon1955class} identified that the process can model the emergence
of power law distributions, citing examples such as
distributions of frequency or words in documents, 
distributions of numbers of papers published by scientists, 
distribution of cities by population, 
distribution of incomes, 
and distribution of biological genera by number of species.
This work was important in the sense that he had discussed small variations
which were put into context. Preferential attachment, sometimes
referred to as \textit{Matthew effect}~\cite{Merton:1968}, is modeled
in various ways, and we will discuss a few processes of historical importance,
which have found prominence in explaining phenomena in nature and society.

The  number of species in a genus, family or other taxonomic group appears to closely 
follow a power law distribution.
Yule~\cite{yule1925mathematical} proposed a model to explain this in the biological context, but is since then, used 
to explain several phenomena, including socio-economic systems. In the following, we will discuss the
model, following Ref.~\cite{newman2005power}.

Let us assume $N$ be the number of genera, which increases by unity with each time step, when
a new species finds a new genus and $m$ other species are added to the already existing
genera  selected in proportion to the number of species they already have. 
Let $P(k,N)$ be the fraction of genera having $k$ species, which makes the number of genera
$NP(k,N)$.
The probability that the next species added to the system adds to 
a genus $i$ having exactly $k_i$ species is proportional to $k_i$, and is $\frac{k_i}{\sum_i k_i} = \frac{mk_i}{N(m+1)}$.
Now, in the interval between $N$th and the $(N + 1)$th genera, 
$m$ other new species are added.
The probability that the genus $i$ gains a new species during this time is simply 
$\frac{mk_i}{N(m+1)}$, while the total number of genera of size $k$ that gain a new species 
during this same interval is $\frac{mk_i}{N(m+1)}NP(k,N) = \frac{mk_i}{m+1}P(k,N)$.
As a result of this evolutionary process, 
(i) there is a decrease in number of genera with $k$ species 
at each time step by this number as they become genera with $k + 1$  by gaining a new species; 
and (ii) along with an increasing term because of species which had $k-1$ species and now have one extra.
One can now write a master equation
\begin{equation}
 (N+1)P(k,N+1)-NP(k,N) = \frac{m}{m+1}\left[ (k-1)P(k-1,N) - kP(k,N)\right],
\end{equation}
while for the genera of size unity
\begin{equation}
 (N+1)P(1,N+1)-NP(1,N) = 1- \frac{m}{m+1}P(1,N).
\end{equation}
One can easily see that $P(1)=\frac{m+1}{2m+1}$. 
The stationary solution for $P(k)$ is
\begin{equation}
P(k) = \left[ k P(k-1) -(k+1)P(k) \right]\frac{m}{m+1}
\end{equation}
which can be rearranged as
\begin{eqnarray}
P(k) &=& \frac{k-1}{k+1+\frac{1}{m}} P(k-1) \\ \nonumber
&=& \frac{(k-1)(k-2) \ldots 1}{(k+1+\frac{1}{m})(k+\frac{1}{m}) \ldots (3+\frac{1}{m})}
=\left(1 +\frac{1}{m}\right)B\left(k,2+\frac{1}{m}\right),
\end{eqnarray}
where $B(x,y)$ is the Beta function. For large $x$ and fixed $y$, $B(x,y) \sim x^{-y}$.
Using this, one gets $P(k) \sim  k^{-(2+1/m)}$.

In the following, we will discuss in brief two closely related models of preferential attachment,
which address the growth of citations to articles. Although the following fit ideally to the 
Sections \ref{sec:biblio} and \ref{sec:networks}, it is best to discuss it at this juncture 
in context of their broad applicability.

The first, due to Price~\cite{Price:1976}, was proposed
in the context of citations is very similar to the Yule process.
It is a model for a directed graph,
and uses the fact that citation networks have
a broad distribution of incoming citations (in-degree)~\cite{Price:1965}. The main idea to exploit is that popular papers
become popular due to \textit{cumulative advantage}~\cite{simon1955class}. 
Let $m=\sum_k k P(k)$ be the mean degree, where $P(k)$ is the degree distribution. 
Price assumes that the  probability that a newly appearing paper cites an older paper is proportional
to the number of citations $k$ of that paper, and in fact taken proportional to $k+1$ to take care of the case of first citation.
Thus,  the probability that a new connection happens on any of the vertices with in-degree $k$ is
\begin{equation}
 \frac{(k+1)P(k)}{\sum_k (k+1)P(k)} = \frac{(k+1)P(k)}{m+1},
\end{equation}
and the average number of new edges (per vertex added) attached to vertices of in-degree 
$k$ will be $(k+1)P(k)m/(m+1)$.
Given $N$ vertices, let $P(k,N)$ denote the probability distribution of in-degree $k$. Thus one can write the master equation as
\begin{equation}
 (N+1)P(k,N+1) - N P(k,N) = \left[ k P(k-1,N) - (k+1) P(k,N) \right] \frac{m}{m+1}, {\rm for}\; k \ge 1,
\end{equation}
with
\begin{equation}
 (N+1)P(0,N+1) -NP(0,N) = 1-P(0,N) \frac{m}{m+1}, {\rm for}\; k =0.
\end{equation}
The stationary solution for $P(k)$ is
\begin{equation}
P(k) = 
\left\{ \begin{array}{lc}
\left[ k P(k-1) -(k+1)P(k) \right]\frac{m}{m+1} & \textrm{for} \ k \ge 1,\\
1-P(0)\frac{m}{m+1} \ \ \ \ & \textrm{for} \  k=0.
\end{array}\right.
\end{equation}
Thus one gets $P(0)=\frac{m+1}{2m+1}$ and $P(k)=\frac{kP(k-1)}{k+2+1/m}$, which leads to
\begin{equation}
 P(k)= \frac{k(k-1)\ldots 1}{(k+2+\frac{1}{m})\ldots(3+\frac{1}{m})}P(0) = \left(1+\frac{1}{m}\right) B\left(k+1,2+\frac{1}{m}\right),
\end{equation}
Using the property of Beta function, 
 $P(k) \sim  k^{-(2+1/m)}$,
which also comes as a result of the Yule process.
The power law exponent in the Price model is always greater than $2$.

The second model, known as the Barab\'{a}si-Albert (BA) model~\cite{barabasi1999emergence}, uses \textit{preferential attachment}
explicitly in its growth dynamics. The main difference from Price's model is that the graph is treated
as undirected, which eliminates the problem with the uncited papers. At each step a new vertex arrives with 
$m (\ge 1)$ edges. Thus, the 
probability that a new edge attaches to any of the vertices with in-degree $k$ is given by
\begin{equation}
 \frac{k P(k)}{\sum_k k P(k)} = \frac{kP(k)}{2m}.
\end{equation}
The average number of vertices of degree $k$ who gain an edge upon adding a single new vertex with $m$ edges is
simply $m k P(k)/2m = kP(k)/2$. Thus, the master equation is
\begin{equation}
 (N+1) P(k,N+1) - NP(k,N) = \frac{1}{2} (k-1) P(k-1,N) - \frac{1}{2} k P(k,N), {\rm for} \; k>m
\end{equation}
and
\begin{equation}
 (N+1)P(m,N+1)-NP(m,N) = 1-\frac{1}{2} mP(m,N), {\rm for} \; k=m.
\end{equation}
The stationary solution for $P(k)$ is
\begin{equation}
P(k) = 
\left\{ \begin{array}{lc}
\frac{1}{2}(k-1)P(k-1) - \frac{1}{2}k P(k)  & \textrm{for} \ k > m,\\
1-\frac{1}{2}mP(m) \ \ \ \ & \textrm{for} \  k=m.
\end{array}\right.
\end{equation}
Thus one gets $P(m) = \frac{2}{m+2}$ and $P(k)= \frac{(k-1)P(k-1)}{k+2}$, combining which one can get
\begin{equation}
 P(k) = \frac{(k-1)(k-2)\ldots m}{(k+2)(k+1) \ldots (m+3)}P(m) = \frac{2m(m+1)}{(k+2)(k+1)k},
\end{equation}
which for large $k$ gives $P(k) \sim k^{-3}$~\cite{Albert:2002,dorogovtsev2002evolution}.

\section{Income, wealth \& energy}
\label{sec:income}
\begin{quotation}
\textit{The disposition to admire, and almost to worship, the rich and the powerful, and to despise, or, at least, to neglect persons of poor and mean condition is the great and most universal cause of the corruption of our moral sentiments.}
\begin{flushright}
-- Adam Smith, Scottish political economist (1723-1790)
\end{flushright}
\end{quotation}
The issue of inequality in terms of income and wealth, is a widely debated subject in economics. 
Economists and philosophers have pondered over the normative aspects of this problem~\cite{Sen-1999,Foucault-2003,Scruton-1985,Rawls-1971}. 
The direct and indirect effects of inequality on the society have also been studied extensively
while relatively less emphasis had been put on the origin of the problem. 
The main non-trivial issues and open questions have been:
related to the form of the income and wealth distributed, 
do they depend upon specific conditions of a country, or if they
are universal, and why, if they are so.

More than a century ago, 
Pareto~\cite{Pareto-book} made extensive studies in Europe and found that wealth distribution follows a power law 
tail for the richer section of the society, known to be the Pareto law. 
Independently, Gibrat worked on the same problem and he proposed a ``law of proportionate effect''~\cite{Gibrat:1931}. 
Much later, Champernowne's systematic approach provided a probabilistic theory to justify Pareto's claim~\cite{Champernowne-1998}. 
Subsequent studies revealed that the distributions of income and wealth possess some globally stable 
and robust features (see, e.g., \cite{chakrabarti2013econophysics}). 
In general, the bulk of the distribution of both income and wealth seems to reasonably fit both the log-normal 
and the Gamma distributions. Economists have a preference for the log-normal distribution~\cite{Montroll:1982,gini1921measurement}, while 
statisticians~\cite{Hogg-2007} and rather recently, physicists~\cite{Chatterjee:EWD,Chatterjee-2007,Yakovenko:RMP} 
prefer the Gamma distribution for the probability density or Gibbs/ exponential distribution for the corresponding cumulative distribution. 
The high end of the distribution, known to be the `tail', is well described by a power law as was found by Pareto.
\begin{figure}[t]
\includegraphics[width=15.0cm]{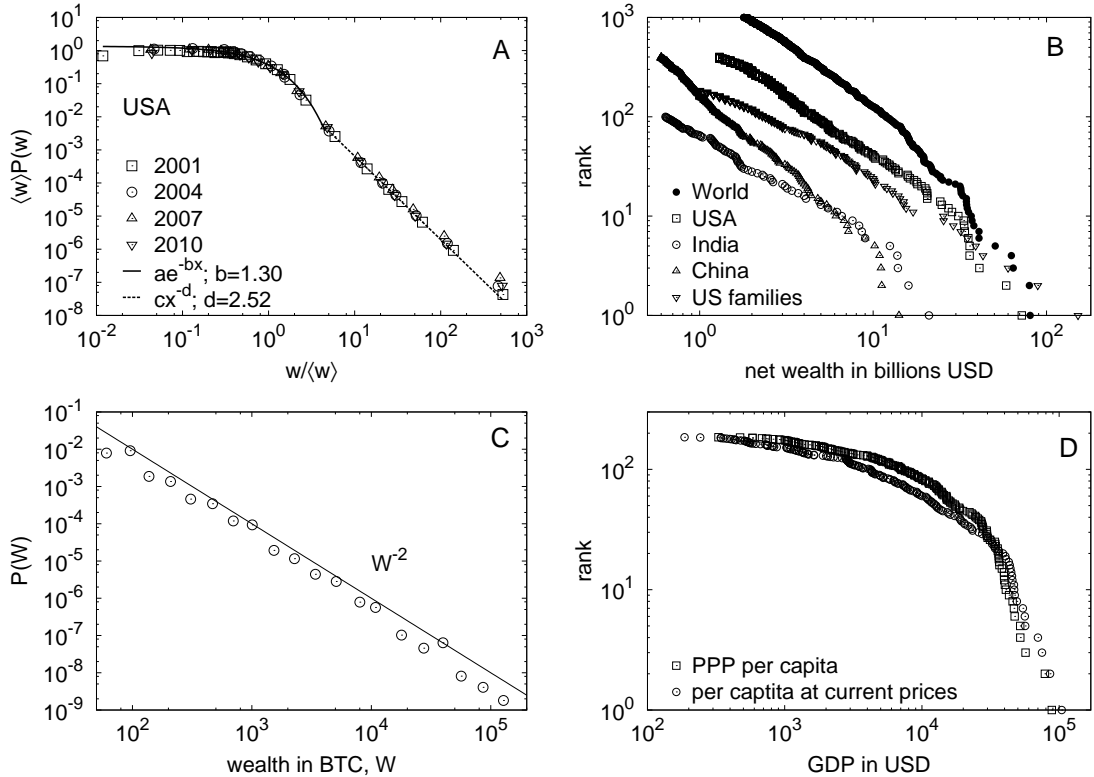}
\caption{
(A) Probability distribution $P(w)$ of income $w$ rescaled by average income $\langle w \rangle$ for USA for different years. 
The income is calculated from the IRS tax data~\cite{IRS}. 
The low and middle income range fits to an exponential $a\exp(-bx)$
while the high income range fits to a power law decay $cx^{-d}$.
(B) The rank plot of the wealthiest individuals, for the whole world, USA, India, China, and also wealthiest families of USA~\cite{Forbes};
(C) The probability distribution $P(W)$ of wealth $W$ (in BTC) of Bitcoin accounts~\cite{BTC};
(D) Rank plot of GDP of countries of the world~\cite{IMF}.
}
\label{fig:income_wealth}
\end{figure}
These observed regularities in the income distribution may thus indicate a ``natural'' law of economics:
\begin{equation}
\label{par}
P(x) \sim
\left\{ \begin{array}{lc}
x^{n} \exp(-x/T) & \textrm{for} \ x < x_c,\\
\frac{\alpha x_c^\nu}{ x^{1+\nu}} \ \ \ \ & \textrm{for} \  x \ge x_c,
\end{array}\right.
\end{equation}
 where $ n$ and $\nu$  are two exponents, and $T$ denotes a scaling factor. 
 The exponent $\nu$ is called the Pareto exponent and its value ranges between 1 and 3~\cite{chakrabarti2013econophysics} 
 (See Ref.~\cite{Wileybook2} for a historical account of Pareto's data and those from recent sources). 
 The crossover point $x_c$ is extracted from the numerical fittings. 
 
Gibrat~\cite{Gibrat:1931} clarified that Pareto's law is valid exclusively for the high income range, 
whereas for the middle income range is described by a log-normal probability density (Eq. \ref{eq:lognormal}).
The factor $\beta=1/\sqrt{2\sigma^2}$ is known as Gibrat index, 
which measures the equality of the distribution. 
Empirically, $\beta$ is found to lie between $2$ and $3$~\cite{Souma-2002}.

Fig.~\ref{fig:income_wealth} shows a few examples of income and wealth distributions --
incomes as computed from income tax data sources~\cite{IRS} of USA for several years, 
wealth of top billionaires~\cite{Forbes},
wealth of individual Bitcoin accounts~\cite{BTC} in terms of Bitcoin units (BTC), 
and also GDP of countries~\cite{IMF} in US Dollars, 
both for purchasing power parity (PPP) per capita as well as per capita at current prices.
All of these indicate a power law for the richest.

It has also been seen that the tail of the consumer expenditure also follows power law 
distribution~\cite{mizuno2008pareto,ghosh2011consumer} along with a lognormal bulk, 
in the same way as income distribution does.

\subsection{Modeling income and wealth distributions}
\label{sec:model_wealth}
Gibrat~\cite{Gibrat:1931}  proposed a law of proportionate effect which explained
that a small change in a quantity is independent of the quantity itself.
Thus the distribution of a quantity $dz = dx/x$
should be Gaussian, and thus $x$ is lognormal, explaining why small and middle range of income
should be lognormal.
Champernowne proposed a probabilistic theory to justify Pareto's claim~\cite{Champernowne:1953,Champernowne-1998},
but used a class structure in a multiplicative process setup (discussed in Sec.~\ref{sec:process}).

In case of wealth distribution, the most popular models are chemical kinetics inspired
Lotka-Volterra models~\cite{Levy:1997,Solomon:2002,Richmond:2001},
polymer physics inspired Bouchaud-M\'{e}zard model~\cite{Bouchaud:2000} and
models inspired by kinetic theory~\cite{Dragulescu:2000,Chakraborti:2000,Chatterjee:2004,Chatterjee-2007} 
(see Ref.~\cite{chakrabarti2013econophysics} for detailed examples).
What is quite well understood and established is that two-class structure~\cite{Yakovenko:RMP}
of the income distribution is a result of very different dynamics of the two classes.
The bulk is defined by a process which is at its core nothing but a random 
kinetic exchange~\cite{Chakraborti:2000,Dragulescu:2000} producing a distribution dominated by an exponential function.
The dynamics is nothing but as simple as what we know as the kinetic theory of gases~\cite{saha1958treatise}.
The minimal modifications that one can attempt are to use additive or multiplicative terms.

Inequality creating processes involving uniform retention rates~\cite{angle1986surplus,angle2006inequality}
or savings~\cite{Chakraborti:2000} can only produce Gamma-like distributions.
In these models, the wealth exchanging entity or \textit{agent} randomly exchanges wealth
but retaining a certain fraction (saving propensity) of what they had before each trading process.
These models consider each agent the same,
assigning each with the same value of the `saving propensity' (as in Ref.~\cite{Chakraborti:2000}),
which could not produce a broad distribution of wealth.
What is important to note here is that the richest follow a different dynamic where 
heterogeneity plays a crucial role. 
To get the power law distribution of wealth for the richest, one needed to simply assume that
each agent is different in terms how much fraction of wealth they save in each trading~\cite{Chatterjee:2004},
a very natural ingredient to assume, since it is very likely that agents in a market think very differently
from one another.
In fact, with this very little modification, on can explain the whole range of wealth distribution~\cite{Chatterjee-2007}.
However, these models can show interesting characteristics if the exchange processes and flows
be made very asymmetric, e.g., on directed networks~\cite{chatterjee2009kinetic}.
Numerous variants of these models, their results and analyses find possible applications
in a variety of trading processes~\cite{chakrabarti2013econophysics,pareschi2013interacting}.

Statistical physics tools have helped to formulate these microscopic models, which are simple
enough yet rich with socio-economic ingredients. Toy models help in understanding
the basic mechanism at play, and bring out the crucial elements that shape the emergent distributions
of income and wealth. A variety of models, from zero-intelligence variants to the much complex
agent based models incorporating game theory have been proposed and found to be successful 
in interpreting results~\cite{chakrabarti2013econophysics}.
Simple modeling has been found to be effective in understanding how entropy maximization produce
distributions dominated by exponentials, and also explain the reasons of aggregation
at the high range of wealth, including the emergence of the power law Pareto tail.

A rapid technological development of human society since the industrial
revolution has been dependent on consumption of fossil fuel (coal, oil, and natural gas),
accumulated inside the Earth for billions of years. The standards of living in our modern 
society are primarily determined by the level of per capita energy consumption. It is now well understood
that these fuel reserves will be exhausted in the not-too-distant future. Additionally,
the consumption of fossil fuel releases CO$_2$ into the atmosphere, which is
the major greenhouse gas, affecting climate globally -- a problem posing
great technological and social challenges.
The per capita energy consumption around
the globe has been found to have a huge variation. 
This is a challenge at the geo-socio-political level and complicates the situation
for arriving at a global consensus on how to deal with the energy issues.
This global inequality in energy consumption was characterized~\cite{Banerjee:2010,lawrence2013global},
and explained using the method of entropy maximization.
 
\subsection{Is wealth and income inequality natural?}
One is often left to wonder if inequalities in wealth and income are natural.
It has been shown in terms of models and their dynamics that certain minimal dynamics~\cite{chatterjee2007economic} 
over a completely random exchange process picture and subsequent entropy maximization produces broad distributions.
Piketty~\cite{piketty2014capital} recently argued that inequality in wealth distribution is indeed quite natural.
He specifically pointed out that before the wars of the early 20th century, the stark skewness of wealth distribution
was prevailing as a result of a certain `natural' mechanism. 
The two great World Wars followed by the Great Depression helped in dispersion of wealth
that, sort of brought the extreme inequality under check and gave rise to a sizable middle class.
He argues, analyzing very accurate data, that the world is currently `recovering' back to its natural state,
due to capital ownership driven growth of finance~\cite{piketty2014inequality} which has been dominant over
a labor economy, which is simply a result of which type of institution and policies are adopted by the society.
This work raises issues, quite  fundamental, concerning both economic theory and the future of capitalism,
and points out large increases in the wealth/output ratio. In standard economic theory, such increases  are associated 
with a decrease in the return to capital and an increase in the wages. 
However, today the return to capital has not been seen to have diminished, while wages have.
There is also some prescription proposed -- higher capital-gains and inheritance taxes, 
bigger spending for access to education, tight enforcement of anti-trust laws, 
reforms in corporate-governance that restrict pay for executives, and finally 
financial regulations which had been an instrument for banks to exploit the society -- all of these 
might help reduce inequality and increase equality of opportunity. The speculation is that
this might be able to restore the shared and quick economic growth that was the characteristic the middle-class 
societies in the middle of the twentieth century.

\section{Urban and similar systems}
\label{sec:urban}
With growing population all over the world, towns and cities have also grown in size (population)~\cite{batty2008size}.
The definitions using  precise delimiters can be an issue of debate, whether one
is looking at a metropolitan area, a commune, or a central urban agglomeration, but irrespective
of that it has been persistently observed that the number of agglomerations of a certain size (or size range)
is inversely proportional to the size. Questions has been asked whether this is a result of 
just a random process, a hierarchical organization, or if they are 
guided by physical and social constraints like optimization or organization.

\subsection{City size}
\begin{figure}[t]
\includegraphics[width=8.0cm]{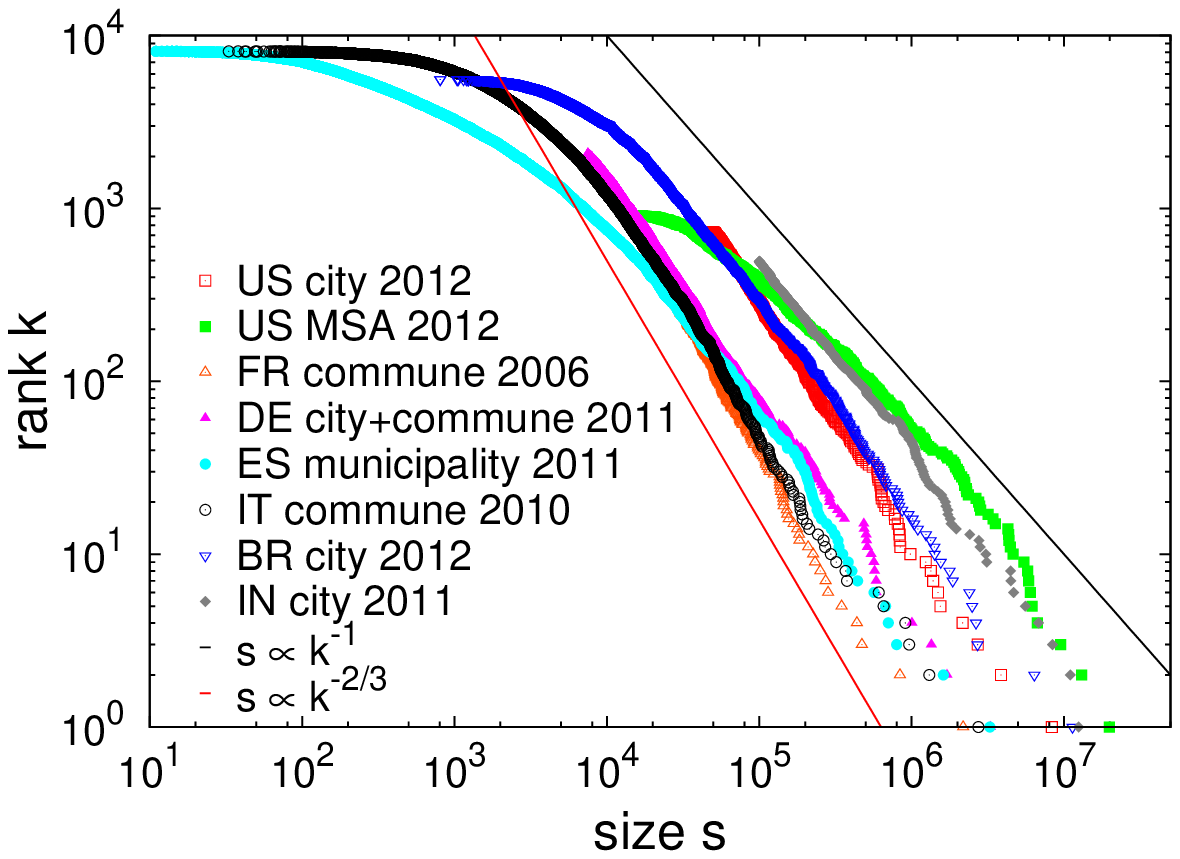}
\includegraphics[width=8.0cm]{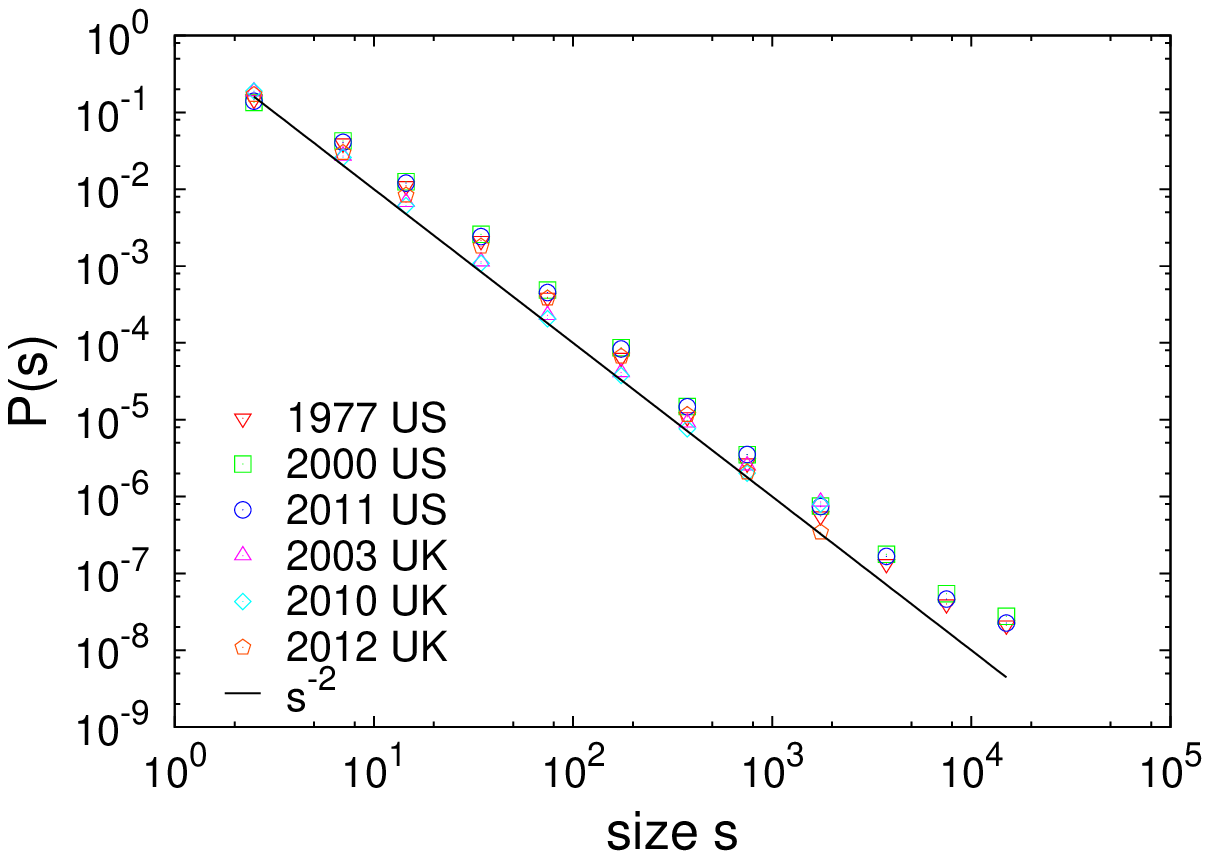}
\caption{Left panel: Plot of size s with rank k of a city for different countries. 
The two straight lines are respectively $s \propto k^{-1}$
and $s \propto k^{-2/3}$, guides to the approximate extremes of the
power law behavior of the data sets.
Figure adapted from Ref.~\cite{ghosh2014city}.
Right panel: Firm size distribution for USA~\cite{US_firm_data} and UK~\cite{UK_firm_data}.
}
\label{fig:city_firm}
\end{figure}

Auerbach~\cite{auerbach1913} was the first to note that the sizes of the cities follow a broad distribution,
which was later bettered by Lotka~\cite{lotka1925elements} as $k^{0.93} N_k = const.$,
which was subsequently cast as~\cite{singer1936courbe} as $kN_k^{\gamma -1} = const.$,
where $N_k$ is the population of the city of rank $k$. Rewriting, one gets 
$k \propto 1/N_k^{\gamma-1}$, where $k$ is the number of cities with population $N_k$ or more.
Thus $\gamma$ is nothing but the size exponent (Pareto).
It was examined again by Zipf~\cite{Zipf:1949} by plotting rank-size distribution of quantities,
and restated as $N_k \propto k^{-b}$, from where it follows that the two exponents are related as 
$1/b=\gamma -1$. $b=1$ is the value of the exponent, \`{a} la Zipf.
Rigorous studies~\cite{rosen1980size} showed that the Zipf exponent has deviations from unity, 
as also found in recent studies of China~\cite{Gangopadhyay2009} or former USSR countries~\cite{Benguigui2007}.
A general view of the broad distribution of city sizes across the world is shown in Fig.~\ref{fig:city_firm} (left panel),
also showing the non-universal nature of the Zipf exponent.

Several studies have attempted to derive Zipf's law  theoretically for city-size distributions,
specifically where the exponent of the CDF of size is equal to unity. 
Gabaix~\cite{gabaix1999zipf1} indicated that if cities grow randomly with the same expected growth rate
and the same variance (Gibrat's law, see Sec.~\ref{sec:income}), the limiting distribution will converge to Zipf's 
law.
Gabaix proposed that growth `shocks' are random and impact utilities both positively and negatively.
In a similar approach of shocks, citizens were assumed to experience them, resulting in 
diffusion and multiplicative processes, producing intermittent spatiotemporal structures~\cite{zanette1997role}.
Another study used shocks as a result of migration~\cite{marsili1998interacting}.
In Ref~\cite{gabaix1999zipf1} however, differential population growth resulted from migration.
Some simple economics arguments showed that expected urban
growth rates were identical across city sizes and variations were random normal deviates, 
and the Zipf law with exponent unity follows.
However, there seemed to be a missing assumption which may not produce the power law~\cite{blank2000power},
a random growth model was proposed based on more `physical' principles, which can generate a power law instead.

\subsection{Scaling of urban measures}
Analysis of available large urban data sets across decades for U.S.A. have concluded that 
(i) size is the major determinant, (ii) the space required per capita shrinks -- the denser the settlement, the more intense is the use of infrastructure,
(iii) all socio-economic activities accelerate, indicating higher productivity, and 
(iv) socio-economic activities diversify and become more interdependent.
What comes as a surprise is that as city size increases, several quantities increase by small factor more than 
linear (\textit{superlinear scaling})~\cite{bettencourt2007growth,bettencourt2010unified}.
These relations were tested to be robust across a variety of urban measures, e.g., crime~\cite{alves2013distance}.
A theoretical framework was developed~\cite{bettencourt2013origins} that derived the general properties of cities
through the optimization of a set of local conditions, and was used
to predict the average social, spatial and infrastructural properties of cities
in a unified and quantitative way. A set of scaling relations were found that apply to
all urban systems, supported by empirical data.

\subsection{Firms}
Sizes of employment firms in terms of employees are known to follow the Zipf law~\cite{axtell2001zipf}, in a
rather robust fashion compared to city sizes. Fig.~\ref{fig:city_firm} (right panel) shows the typical
size distribution features in case of firms in USA and UK.

However, recent studies find that while the Zipf law
holds good for the intermediate range, if one considers the firm assets, the power law which is usually attributed to 
a mechanism of proportional random growth, based on a regime of constant returns to scale.
However the power law breaks down for the largest firms, which are mostly financial firms~\cite{fiaschi2014interrupted}. 
This deviation from the expected size, \`{a} la Zipf law, has been attributed to the \textit{shadow banking system},
which can broadly be described as credit intermediation involving entities and activities
outside the regular banking system. The identification of the correlation of the size of the
shadow banking system with the performance of the financial markets through crashes and
booms sheds some light on the reason for the financial crises that have struck the world in
the recent times.

\section{Consensus}
\label{sec:consensus}
\begin{quotation}
 \textit{Consensus is what many people say in chorus but do not believe as individuals.}
\begin{flushright}
  -- Abba Eban, Israeli politician (1915-2002)
\end{flushright}
 \end{quotation}
Consensus in social systems is a very interesting subject in terms of it dynamics,
as well as concerning conditions under which it can happen.
Consensus is one of the most important aspects of social dynamics. 
Life presents many situations where it  requires to assess shared decisions. 
Agreement leads to stronger position, giving it a chance to have an impact on society.
The dynamics of agreement and disagreement among
a collection of social being is complex. Statistical physicists working on opinion dynamics
tend to define  the opinion states of a population and the dynamics that determine the transitions between
such states. 
In a typical scenario, one has  several opinions (discrete or continuous) and one studies 
how global consensus (i.e., agreement of
opinions) emerges out of individual opinions~\cite{liggett1999stochastic,deffuant2000mixing,hegselmann2002opinion}.
Apart from the dynamics, the interest lies in the distinct steady state properties: a phase characterized
by individuals widely varying in  opinions and another phase where the majority of individuals have similar opinions.

\subsection{Voting}
\begin{figure}[t]
\begin{center}
\includegraphics[width=14.0cm]{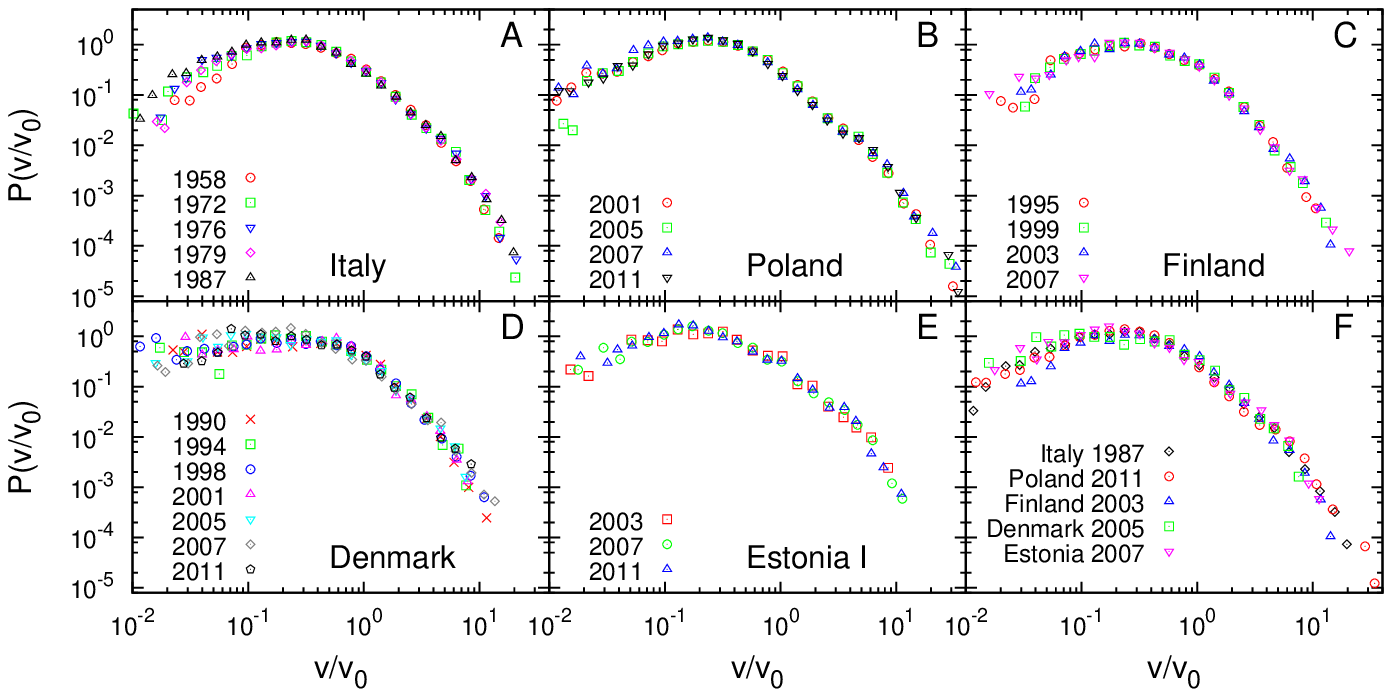}
\vskip -0.5cm
\includegraphics[width=9.0cm]{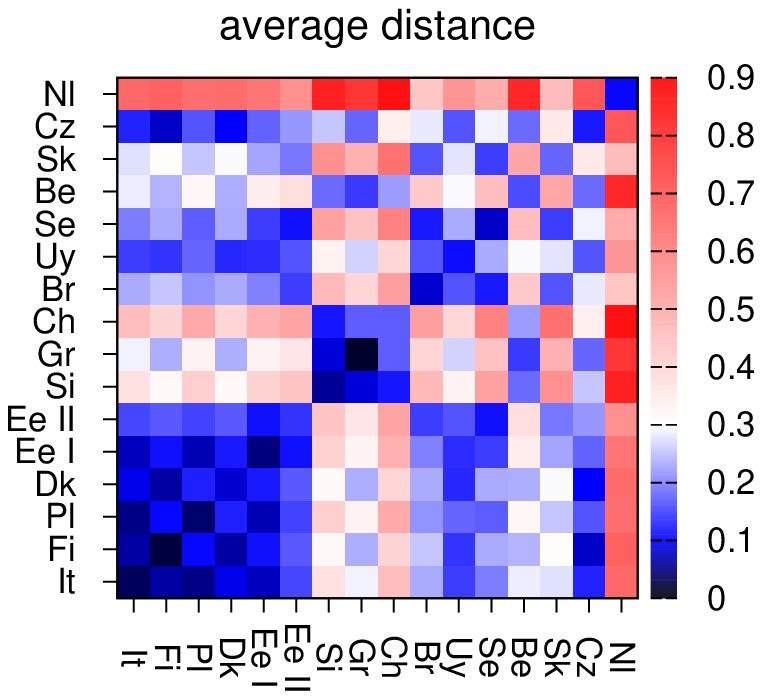}
\vskip -0.5cm
\includegraphics[width=7.0cm]{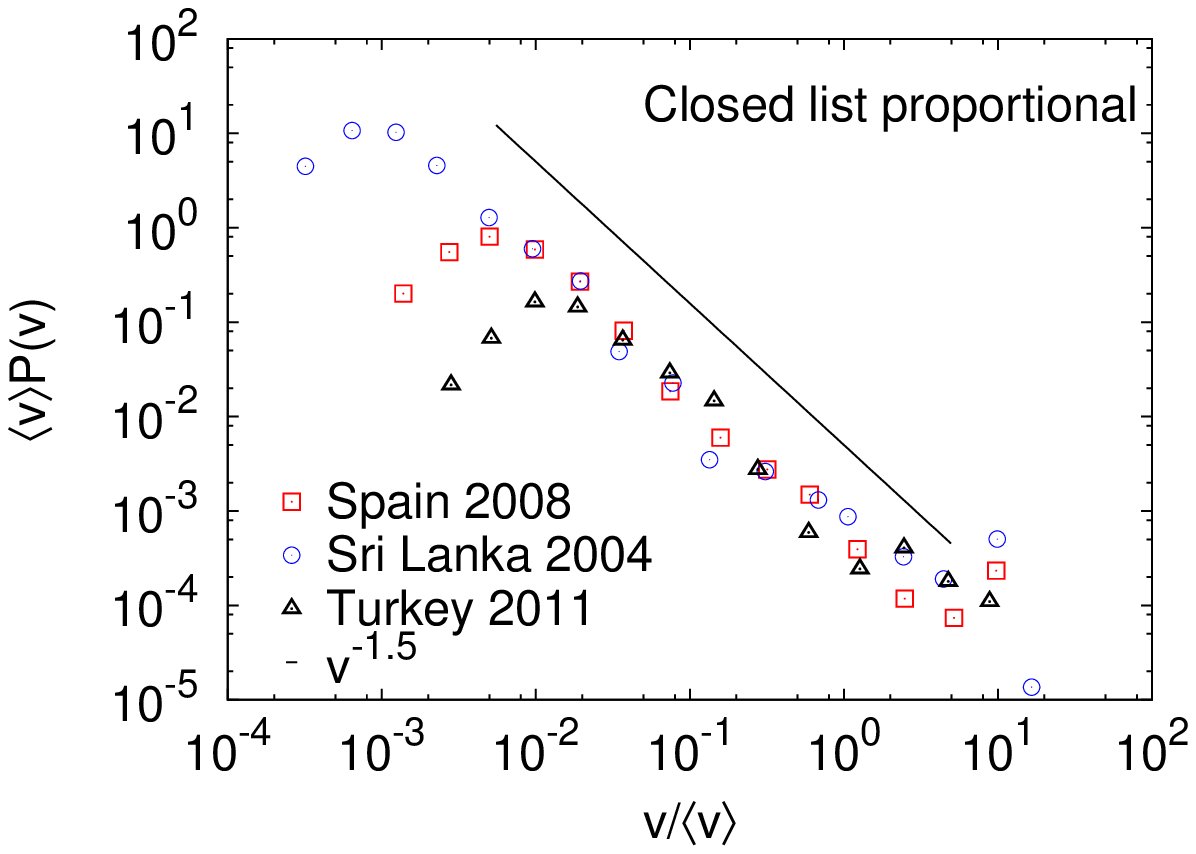}
\includegraphics[width=7.0cm]{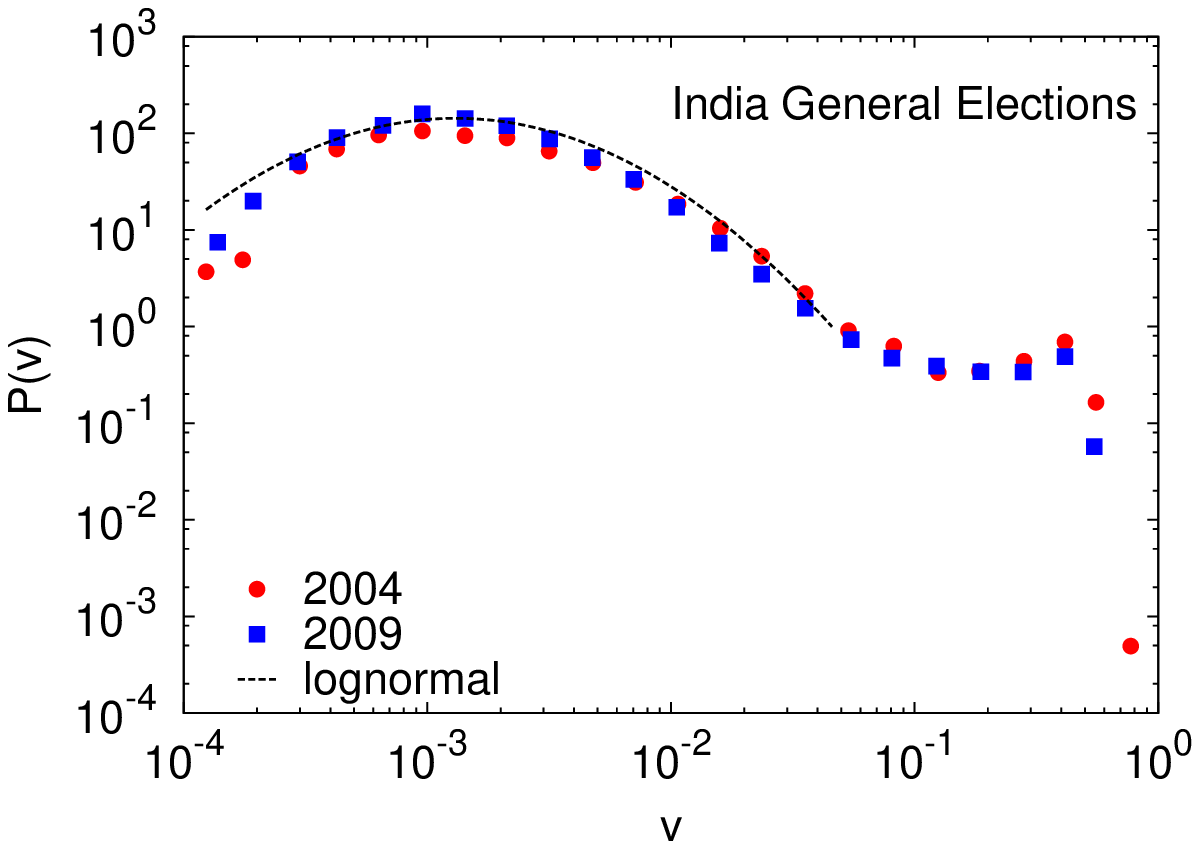}
\end{center}
\caption{
Top: Distribution of electoral performance of candidates in proportional elections with open lists.
Middle: Quantitative assessment of the similarity between distributions between countries --
The average K-S distance between distributions of electoral performance of candidates.
Figures are adapted from Ref.~\cite{chatterjee2013universality}.
Bottom Left: Probability distribution $P(v)$ of fraction of votes $v$ won by lists, rescaled
by the average $\langle v \rangle$ for Closed List Proportional voting systems. The solid line 
indicates a power law decay of $v^{-1.5}$.
Bottom Right: Probability distribution $P(v)$ of fraction of votes $v$ won by candidates
in parliamentary elections of India, which follow majority rule voting.
}
\label{fig:fc}
\end{figure}
The most common example of consensus formation in societies are in the process of voting.
Elections are among the largest social phenomena, and  has been well studied over years.
There are many studies concerning  turnout rates,
detection of election anomalies,
polarization and tactical voting,
relation between party size and temporal correlations,
relation between number of candidates and number of voters,
emergence of third parties in bipartisan systems etc.
Since the electoral system varies from country to country,
there exists debates on the issue of which of the systems
are more effective in capturing the view of the society.
One of the many interesting questions one can ask
is whether the distribution of votes is universal in a particular type of voting system.
Recently it was shown that for open list proportional elections,
the probability distribution of a performance of a candidate with respect to 
its own party list follows an universal lognormal distribution~\cite{fortunato2007scaling}, later
confirmed by an extensive analysis of 53 data sets across 15 
countries~\cite{chatterjee2013universality}.
While the countries
using open list proportional system do show universality, the rest follow different distributions
showing little commonality within their patterns.
In Fig.~\ref{fig:fc} (top panel) we show the data for countries which follow the lognormal
pattern. The easiest way to see if distributions are similar, is to
perform the Kolmogorov-Smirnov (K-S) test, where one can compute a distance measure 
to express how dissimilar a pair of probability distributions are.
Fig.~\ref{fig:fc} (bottom panel) shows these distances.
The countries at the left bottom corner, i.e., Italy, Finland, Poland, Denmark, Estonia I (elections held after 2002)
are similar and their performance distributions known to be lognormal. Data from other countries are less similar to this set,
and hence the computed K-S distances are much larger.

 In  \textit{closed list} proportional system, votes are given to lists rather than candidates.
 Usually, the voting mechanism is complicated, with specific rules adopted by specific countries.
 What one can measure here is the fraction of votes $v$ given to a list. In Fig.~\ref{fig:fc} (Bottom Left),
 we show the rescaled (with average votes $\langle v \rangle$) plots for one election each of Spain, Sri Lanka and Turkey.
 What is interesting to observe is that the basic nature of the curves are similar, the range of the abscissa
 being dependent on the size of the constituencies. For a considerable range, the curves seem to follow a power law,
 approximately $v^{-1.5}$.

The majority rule or the \textit{first-past-the-post} (FPTP) is a simple plurality voting system
where the candidate with the maximum votes is declared the winner, 
and hence referred to as the \textit{winner takes all} system.
Usually, several political parties compete in a constituency, each nominating one candidate for the
election. Hence, in practice, the list of candidates in a constituency consist of one each from 
all political parties contesting against one another.
A voter casts a single vote in favor of the candidate of his/her choice. 
Under the FPTP system, the candidate with the maximum vote wins and is elected to the parliament.
Fig.~\ref{fig:fc} (Bottom right) shows the distribution $P(v)$ of the fraction of votes $v$ received 
by a candidate in its constituency~\cite{FPTPpaper}.
The curves are similar, with a log-normal peak in the bulk, indicative of an effective
random multiplicative/branching process, and a second `leader-board' peak around 
$v^* \approx 1/2$, indicative of the effective competition in a FPTP system.
The gross feature of FPTP elections are similar except that the size and position
of the leader-board peak indicated the effective competition in the political scenario.

Mathematical modeling would gain immensely from the information
obtained from the empirical results.
Having established the regularities for different election scenarios, 
one of the future challenges will be to develop statistical methodologies
which can detect/indicate malpractices or even frauds in elections.

\subsection{Religion}
\begin{figure}[t]
\begin{center}
\includegraphics[width=9.0cm]{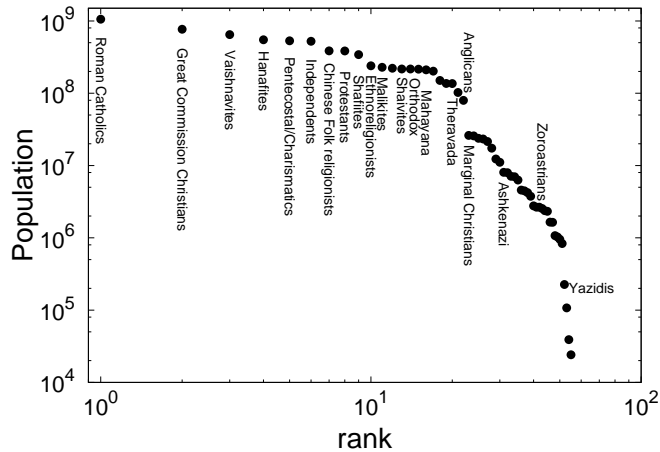}
\end{center}
\caption{
The rank plot of the number of religious adherents -- numbers of different sub groups.
Data taken from Ref.~\cite{rel1}.
}
\label{fig:rel}
\end{figure}

A very debatable question is whether religion can be treated as a process of consensus formation?
Viewing religious ideologies as opinion states, one can easily visualize the dynamics of religion -- as in conversion,
growth and decay in population due to change in the level of popularity, paradigm shift in beliefs, etc.
Fig.~\ref{fig:rel} shows the rank plot of major religious affiliations.
There has already been studies~\cite{ausloos2007statistical} relates to the size distribution
of religious groups. The issue of inequality rises from the fact that religious representations
are very much inhomogeneous across countries, the usual trend being the prevalence of one dominating religion 
or religious ideology. The cultural aspects, which are highly correlated with religious ideology
shapes the social conditions and level of integration between various groups.

\subsection{Modeling opinion and its dynamics}
Statistical physics has been used extensively in modeling opinion and its dynamics~\cite{holyst2002social,san7binary,stauffer2009opinion,sobkowicz2009modelling,weisbuch2006,Castellano:2009}.
The crucial step in modeling opinion is to assign the opinion `states'.
In situations of binary choice, the Ising model paradigm seems to work,
while  situations with multiple choices can be modeled in various ways, e.g. Potts spin model
for discrete case, or a continuous variable defined within a certain range of real numbers.

\subsubsection{Discrete opinion models}
Among the earliest models proposed to describe opinion formation in a society is the voter
model~\cite{clifford1973model,liggett1985interacting}, where opinions can assume two values, 
$0$ and $1$ or $-1$ and $+1$. At each step of dynamics a randomly chosen individual is
assigned the opinion of one of its randomly chosen nearest neighbors, independent of the original opinion
state. The model represents a society where agents only imitate others.
This model has similarities with the Ising model at zero temperature,
where the states of the spins evolve depending to the states of the neighbors. 
But, since a random neighbor is chosen, the probability that any particular opinion is chosen 
is proportional to the number of neighbors having that opinion state,
which is different from the Ising model, in which the probabilities depend exponentially on 
that number. For dimension $d > 2$, no consensus can be reached for infinite systems,
while in finite lattices consensus is asymptotically  reached. However, 
for $d \le 2$, it is possible to reach a consensus.

The Sznajd model~\cite{sznajd2000opinion}  was motivated
by the phenomenon of social validation, i.e., agents are influenced
by groups rather than individuals. In a one-dimensional lattice, a pair of neighboring spins
influence the opinion of their other neighbors in a particular way. In its original version, the
rules of dynamic evolution were defined as:
\begin{itemize}
 \item if $s_is_{i+1} =1$, then $s_{i-1},s_{i+2}$  assumes the opinion of $s_i$
 \item if $s_is_{i+1} =-1$, then $s_{i-1}$  assumes the opinion of $s_{i+1}$, $s_{i+2}$  the opinion of $s_i$.
\end{itemize}

In the majority-rule model~\cite{galam2002minority}
a group of size $R$ is randomly chosen and the opinion state which is the majority in that group is assigned to all these $R$ individuals,
$R$ being  a random variable. One of the `favored' opinions is adopted when there is a tie. 
Depending on the initial configuration, the final state is $+1$ or $-1$ for all opinions.
Let the favored opinion be $+1$.
It was shown that if one starts initially with $p$ fraction of opinions equal to $+1$, 
then a phase transition occurs at $p_c$ -- for $p > p_c$, the final state is all $+1$. 
$p_c$ is dependent on the maximum size of a group. It was shown that $p_c$ could be less than $1/2$, 
which means that an initial minority can win in the end. 
The bias is possibly the reason for this, and was termed `minority spreading'. 
However, if only odd-sized groups are allowed, $p_c = 1/2$. 
This model framework has been used with minor modifications to explain hierarchical voting,
multi-species competition, etc. (see Ref.~\cite{galam2008sociophysics} for details).
The problem with constant (odd) $R$ was solved in the mean-field limit~\cite{krapivsky2003dynamics}, 
where the group can be formed with randomly selected agents. 
The consensus time was found to be exactly $\ln N$ for $N$ agents.
In $d=1$, where the opinions are not conserved, it was reported that the density of domains 
decay according to $t^{-1/2}$ so that the time to consensus is $\mathcal{O}(N^2)$. 
In $d >1$ there is a broad distribution of the consensus times in higher dimensions and the most probable time to reach
consensus shows a power-law dependence on $N$ with an exponent which depends on $d$. 
The upper critical dimension is claimed to be greater than $4$
and in dimensions greater than one there are always two characteristic timescales present in the system.

\subsubsection{Continuous opinion models}
One of the important models using opinion as a continuous variable is the Deffuant model, 
uses the idea of `bounded-confidence'~\cite{deffuant2000mixing}. 
In a model where a pair of individuals simultaneously change their opinions, it was
assumed that two agents interact only if their opinions are close enough,
as in people sharing closer points of view would interact more.
Let $o_i(t)$ be the opinion of the individual $i$ at time $t$. An interaction of the agents $i$ and $j$ agents, 
who are selected randomly, would make the opinions update according to
\begin{equation}
 \begin{aligned}
o_i(t + 1) &=& o_i(t) + \mu[o_j(t) - o_i(t)], \\
o_j(t + 1) &=& o_j(t) + \mu[o_i(t) - o_j(t)],
\end{aligned}
\end{equation}
where $\mu$ is a constant ($0 \le \mu \le 0.5$), known as the `convergence parameter'.
The total opinion remains constant and bounded, i.e., $o_i$ lies in the interval $[0,1]$. 
There is no randomness in the original  model except the random
choice of the agents who interact. Obviously, if $o_j > o_i$,  $o_i$
becomes closer to $o_j$. If $\delta$ is the `confidence level' (agents interact when their
opinions differ by a quantity not more than $\delta$), the final distribution of opinions is dependent on 
on the values of $\delta$ and $\mu$. 
The possibilities are:
(i) all agents may attain a unique opinion value, which is a case of consensus. 
It has been reported that the threshold value of $\delta$ above which all agents have 
the same opinion in the end is $1/2$, and is independent of the underlying topology~\cite{fortunato2004universality}.
(ii) Convergence is reached with only two opinions surviving, a case of `polarization'.
(iii) It can happen that the final state has several possible opinions: a state of `fragmentation'.

Hegselmann and Krause~\cite{hegselmann2002opinion} generalized the Deffuant model,
where an agent simultaneously interacts with all other agents who have opinions within
a prescribed bound. Instead of the convergence parameter, the simple average of the opinions
of the appropriate neighbors is adopted. As a result, consensus is enhanced by making the
bound larger. In case of a graph, the value of the consensus threshold is seen to depend on the average degree.
In these models, starting from continuous values, the final opinions are discretized.

Another class of models employ a kinetic exchange like framework~\cite{toscani2006kinetic}.
Including a diffusion term that takes care of the fact that agents can be randomly affected by external factors
and subsequently change their opinion. The opinions $o_i$ and $o_j$ evolve following
\begin{equation}
\begin{aligned}
o_i(t + 1) &=& o_i(t) + \gamma \mathcal{P}(|o_i|) [(o_j(t)- o_i(t)) + \eta_i \mathcal{D}(|o_i|)],\\
o_j(t + 1) &=& o_j(t) + \gamma \mathcal{P}(|o_j|) [(o_j(t)- o_i(t)) + \eta_j \mathcal{D}(|o_j|)],
\end{aligned}
\end{equation}
where $\gamma \in [0, 1/2]$  represents compromise propensity and $\eta$ is drawn from a random distribution with zero mean. 
As in the bounded-confidence models, the opinion of an agent will tend to decrease or increase so as to be closer to the
opinion of the other. it is important to note that even if the diffusion term is absent,
the total opinion is not conserved unless $P(o)$ is a constant. The functions $\mathcal{P}(.)$ and $\mathcal{D}(.)$ take
care of the local relevance of the compromise and diffusion terms respectively. The choice of the functional forms of $\mathcal{P}$ and $\mathcal{D}$
decides the final state.

Another simpler  model~\cite{lallouache2010opinion}
considers binary interactions, where the opinions evolve as:
\begin{equation}
\begin{aligned}
o_i(t + 1) = \lambda [o_i(t) + \epsilon o_j(t)],\\
o_j(t + 1) = \lambda [o_j(t) + \epsilon^\prime o_i(t)],
\end{aligned}
\end{equation}
where $\epsilon, \epsilon^\prime$ are drawn randomly from uniform distributions in $[0, 1]$. 
$o_i(t)$ is the opinion of individual $i$ at time $t$, and $o_i \in [-1,+1]$, and are also bounded, i.e.,.
if $o_i$ exceeds $1$ or becomes less than $-1$, it is set to $1$ and $-1$, respectively.
The parameter $\lambda$ is interpreted as `conviction', and this models a society in where 
everyone has the same value of conviction. The model lacks conservation laws.
The order parameter is defined as $O = |\sum_i o_i|/N$, the magnitude of the average opinion in a system with N agents. 
One also measures the fraction of the agents having $o_i = \pm 1$, called the condensation fraction $p$
Numerical simulations indicate that the system goes into either of the two possible phases: for
any $\lambda \le \lambda_c$, $o_i = 0$ $\forall i$, 
while for $\lambda > \lambda_c$, $O > 0$ and $O \to 1$ as $\lambda \to 1$ with $\lambda_c \simeq 2/3$,
$\lambda_c$ is the `critical point' of this phase transition. 
The critical point is easily confirmed by mean field calculation~\cite{biswas2011phase}.
Using a fixed point equation $o^* [1-\lambda (1+\langle \epsilon \rangle)]=0$, the fixed point turns out to
be $\lambda_c = 1/(1+\langle \epsilon \rangle)$. For a random uniform distributed $\epsilon$ with $\langle \epsilon \rangle =1/2$,
one can easily see, $\lambda_c=2/3$.

Several variants including the discrete version~\cite{biswas2011mean},
and the generalized version with a second parameter `influence'~\cite{sen2011phase}
gave further insights to the class of models. In another model~\cite{biswas2012disorder},
where negative influence was also considered, one could consider
both discrete and continuous versions, which gave interesting insight into 
competing views and phase transitions.

All of the above models, which show a phase transition, can be viewed to
correspond to broad distributions of opinions/consensus variables
near their critical point. It is much easier to visualize a percolation~\cite{stauffer1991introduction}
picture, where the sizes of percolation clusters have a power-law distribution close to
the percolation phase transition. The alternative picture is that of self-organized criticality~\cite{bak1997nature},
where the activity size is also power law distributed.

\section{Bibliometrics}
\label{sec:biblio}
Academic publications (papers, books etc.) form an unique social system consisting
of individual publications as entities, containing bibliographic reference to other older publications,
and this is commonly termed as \textit{citation}. 
The number of citations is a measure of the importance of a publication, and serve as a proxy
for the popularity and quality of a publication.
There has already been a plethora of empirical studies on citation data~\cite{Sen:2013}, specifically on 
citation distributions~\cite{Shockley:1957,Laherrere:1998,Redner:1998,Radicchi:2008} of articles,  
time evolution of probability distribution of citation~\cite{Rousseau:1994,Egghe:2000,Burrell:2002}, 
citations for individuals~\cite{Petersen:2011} and even their dynamics~\cite{Eom:2011},
and the modeling efforts on the growth and structure of citation networks have produced a huge body literature
in network science concerning scale-free networks~\cite{Price:1965,barabasi1999emergence,caldarelli2007scale}.

The bibliometric tool of citation analysis is becoming increasingly popular
for evaluating the performance of individuals, research groups, institutions as well as countries,
the outcomes of which are becoming important in case of offering grants and awards, 
academic promotions and ranking, as well as jobs in academia, industry and otherwise.
Since citations serve as a crucial measure for the importance and impact of a research
publication, its precise analysis is extremely important.
It is quite usual to find that some publications do better than others
due to the inherent heterogeneity in the quality of their content, the gross attention 
on the field of research, the relevance to future work and so on. Thus different publications 
gather citations in time at different rates and result in a broad distribution of citations.
For individual publications, the present consensus is that while most part of the distribution 
fit to a lognormal~\cite{Shockley:1957,Redner:2005}, 
the extreme tail fits to a power law close to $3$~\cite{Redner:1998,peterson2010nonuniversal}.
This inequality in citations and other associated bibliometric indicators
has been a field of interest to scientists in the last few decades. 

There has also been  studies that address the most popular papers, most cited scientists,
and even Nobel prizes, which are awarded for groundbreaking discoveries, which change the face of science for ever.
The laureates belong to the elite, and the papers which are identified to be the ones
declaring the `discovery' are timeless classics by their own right.
Speaking in terms of \textit{inequality}, these discoveries are indeed rare, and belong to the extreme 
end of the spectrum of all the associated work done in a field, some leading to the discoveries themselves, 
others complementing them,  naturally true for the Nobel prize winning papers.
However, there are limitations to the prize itself, defined by the number of discoveries that can be recognized and 
number of recipients in a single year in a given discipline.
It has been observed that the delay between the discovery and the recognition as a Nobel prize is growing 
exponentially~\cite{fortunato2014growing,becattini2014nobel}, the most rapidly in Physics, followed by Chemistry and 
Physiology or Medicine. As a result the age of a laureate at the time of the award is also increasing, seen to be exponential.
Comparison with the average life expectancy of individuals  concluded that by the end of this century, 
potential Nobel laureates in physics and chemistry are most likely to expire before being awarded and recognized.

In recent times, big team projects have dominated some of the frontline areas
of science, in astrophysics, biology, genetics, quantum information etc.
One can easily see that there has been a gorss inequality in the number of 
researchers who coauthor a paper~\cite{milojevic2014principles}.

In what follows, we will limit our discussion to some interesting aspects related to analysis of citations.

\subsection{Annual citation indices}
One of the quantities of major interest is the nature of the tail of the distribution
of annual citations (AC) and impact factor (IF). 
IF are calculated annually for journals which are indexed in the Journal Citation Reports~\cite{JCR}.
The precise definition is the following: if papers published in a journal
in years $T-2$ and $T-1$ are cited ${\mathcal {N}}(T-2) + {\mathcal {N}}(T-1)$ times by the indexed journals in the year $T$, 
and $N(T-2) + N(T-1)$ are the number of `citable' articles published in these years, 
the impact factor in year $T$ is defined as
\begin{equation}
\label{eq:if}
I(T) = \frac{{\mathcal {N}}(T-2) + {\mathcal {N}}(T-1)}{N(T-2) + N(T-1)}.
\end{equation}
The  number of annual citations (AC) $n(T)$  to a journal in a given year is 
\begin{equation}
\label{eq:cita}
n(T) = \sum_{t \leq T}{\sum_i {\mathcal{A}}_i (t,T)},
\end{equation}
where ${\mathcal {A}}_i(t,T)$ is the number of citations received in the year $T$ 
by the $i$th paper published in the year $t \leq T$.

Another measure, $r(T)$, the \textit{annual citation rate} (CR)
  at a particular year $T$ 
that is   defined~\cite{khaleque2014evolution} as annual citations
divided by the number of articles published in the same year, i.e.,
\begin{equation}
\label{eq:cita-rate}
r(T) = n(T)/N(T).
\end{equation}

\begin{figure*}
 \begin{center}
 \includegraphics[width=11.9cm]{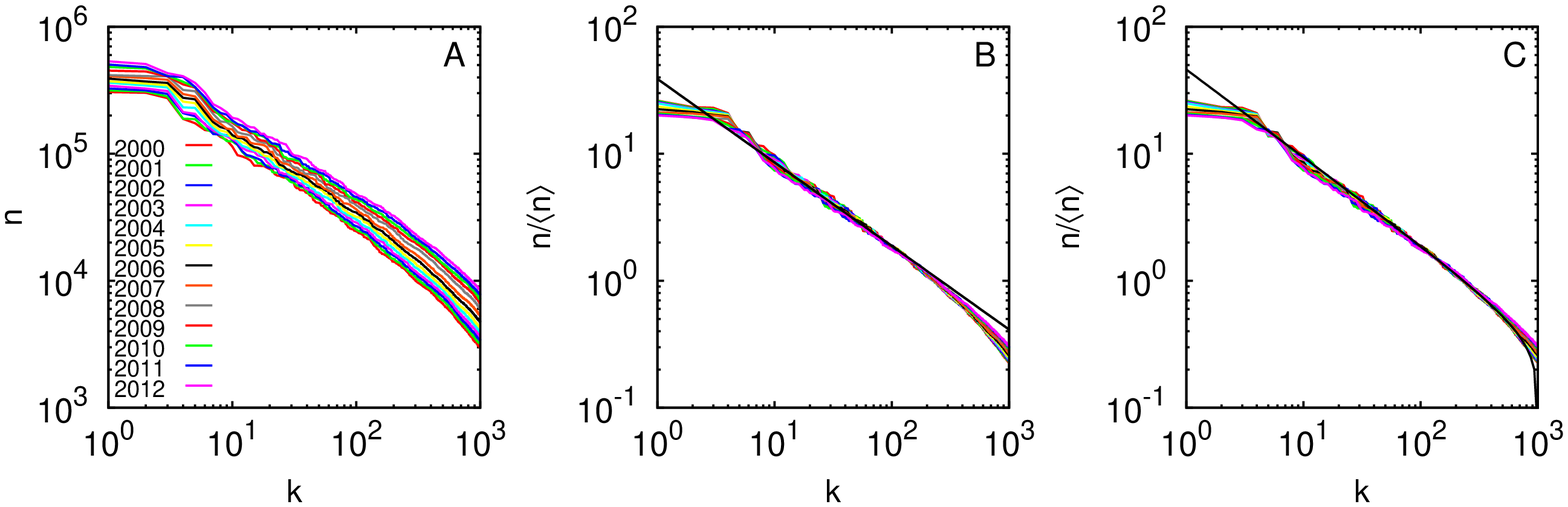}
 \vskip -0.1cm
  \includegraphics[width=11.9cm]{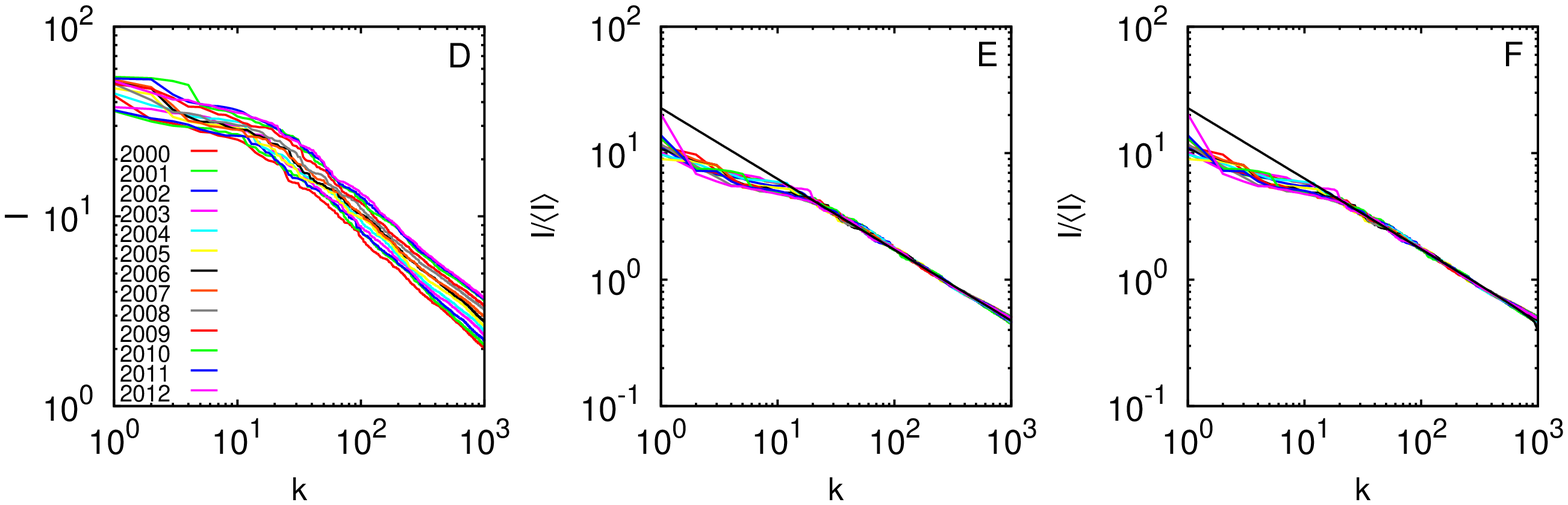}
 \vskip -0.1cm
 \includegraphics[width=11.9cm]{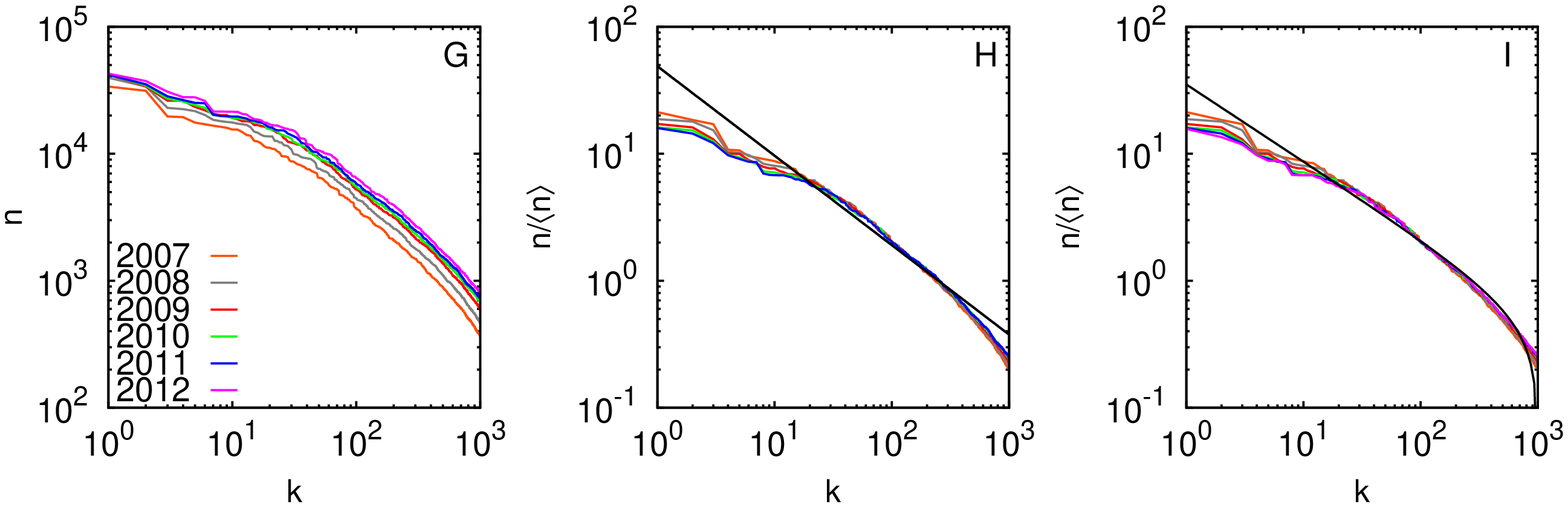}
 \vskip -0.1cm
 \includegraphics[width=11.9cm]{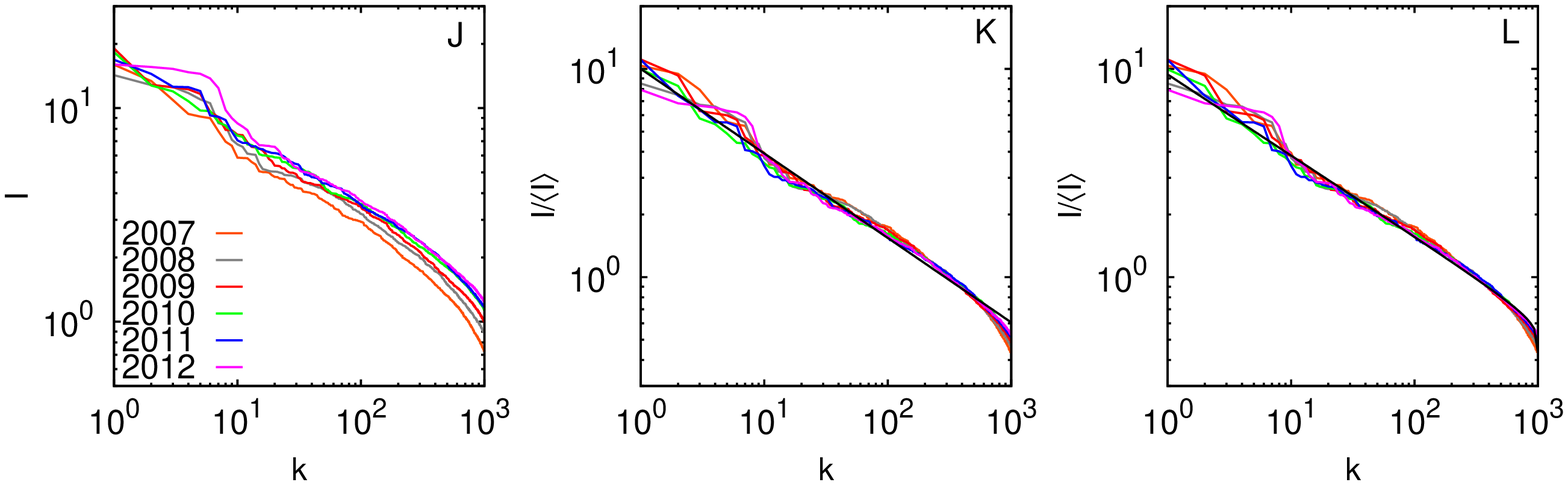}
    \end{center}
   \caption{Plots of annual citation $n$ and impact factor $I$ with rank $k$: 
   (A) Rank plot of the top $1000$ journals, ranked according to citations,
    scaling collapse of the same 
    (B) showing a Zipf law fit: $f(x)=Ax^{-b_n}$, with 
$b_n = 0.70(2)$ and
(C) showing a two-exponent Zipf law fit: $f(x)=k(N+1-x)^{a_n}/x^{b_n}$, with $a_n = 0.27(2)$ and 
$b_n = 0.69 (2)$.
    (D) rank plot of the top $1000$ journals, ranked according to impact factors, and
 scaling collapse of the same (E)  showing a Zipf law fit: $g(x)=Ax^{-b_I}$, 
with $b_I=0.54(1)$ and
(F)  showing a two-exponent Zipf law fit: $f(x)=k(N+1-x)^{a_I}/x^{b_I}$, with $a_I = 0.03(2)$ and 
$b_I= 0.56 (2)$.
 The data is from SCI sets.
 SOCSCI data: (G) Rank plot of the top $1000$ journals, ranked according to citations,
    scaling collapse of the same (H) showing a Zipf law fit: $f(x)=Ax^{-b_n}$, 
with  $b_n=0.7043 \pm 0.001$ and 
(I) showing a two-exponent Zipf law fit: $f(x)=k(N+1-x)^{a_n}/x^{b_n}$, with $a_n = 0.47(2)$ and 
$b_n = 0.61 (2)$.
    (J) rank plot of the top $1000$ journals, ranked according to their impact factors, 
 scaling collapse of the same (K) showing a Zipf law fit: $f(x)=Ax^{-b_I}$, with $b_I= 0.40(1)$ and 
(L)  showing a two-exponent Zipf law fit: $f(x)=k(N+1-x)^{a_I}/x^{b_I}$, with $a_I = 0.06(2)$ and 
$b_I= 0.39 (2)$. The figure is adapted from Ref.~\cite{khaleque2014evolution}.
}
\label{fig:rank}
\end{figure*}
\begin{figure}
\begin{center}
\includegraphics[width=9.5cm]{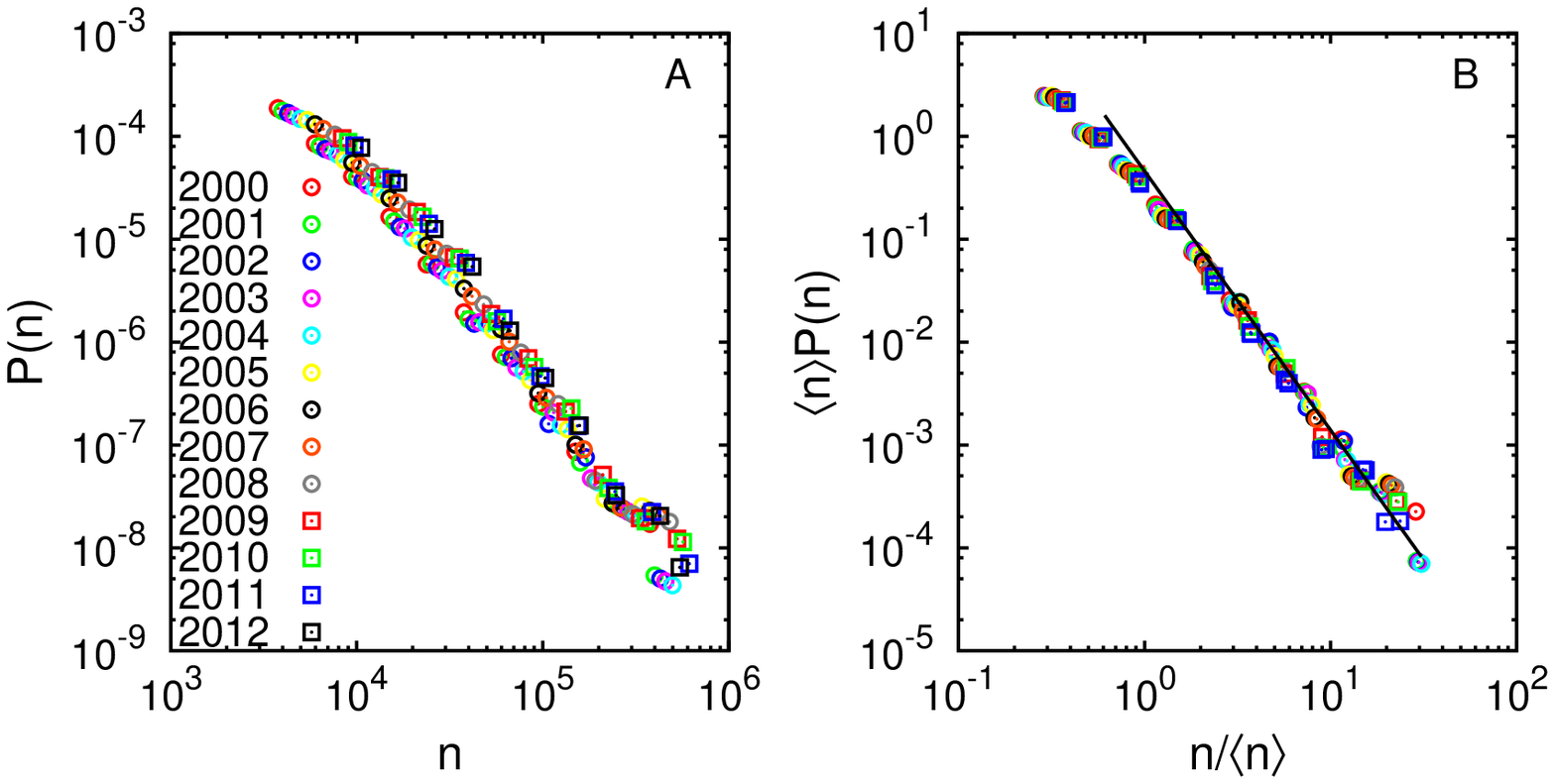}
\includegraphics[width=9.5cm]{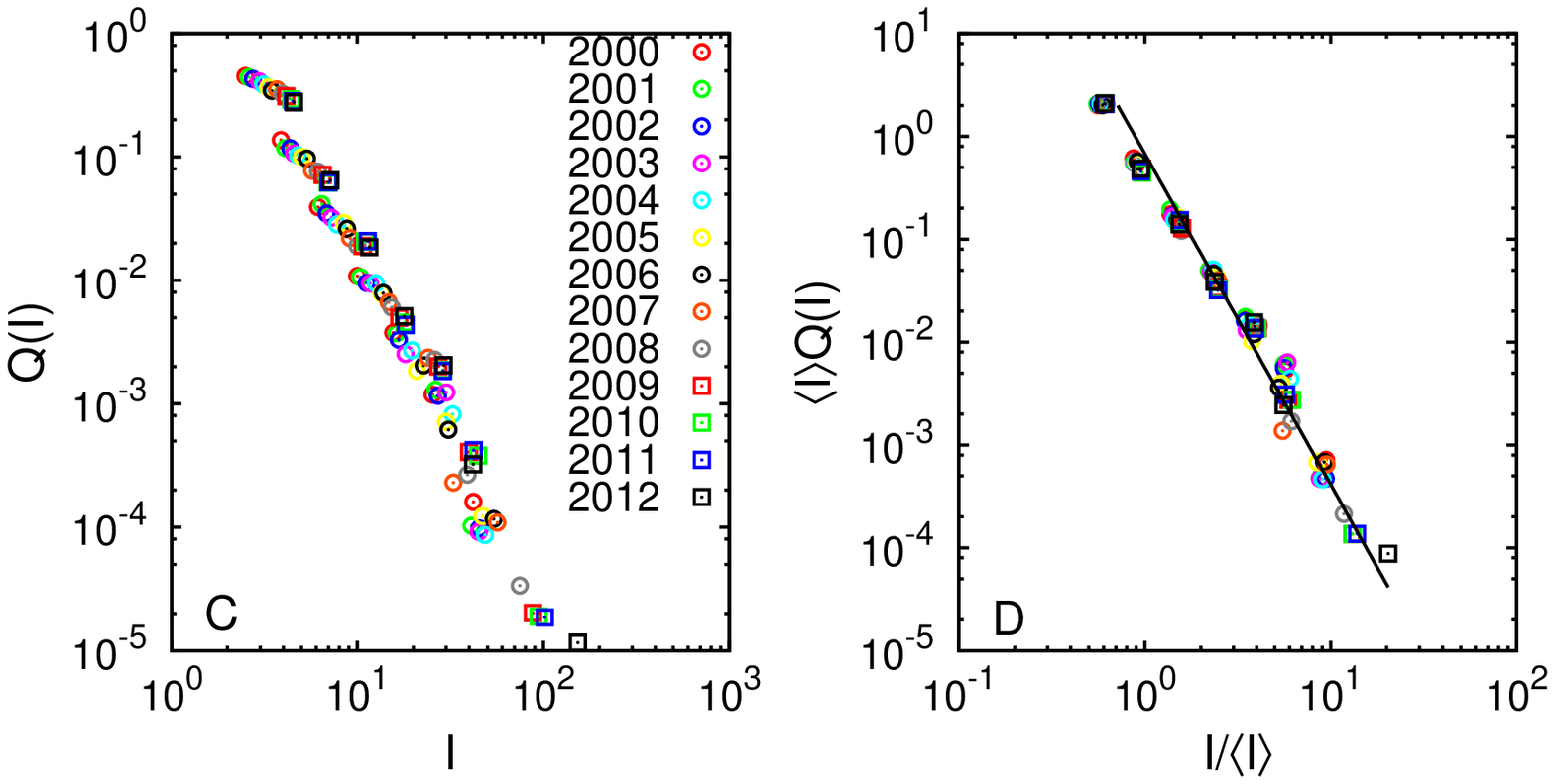}
\includegraphics[width=9.5cm]{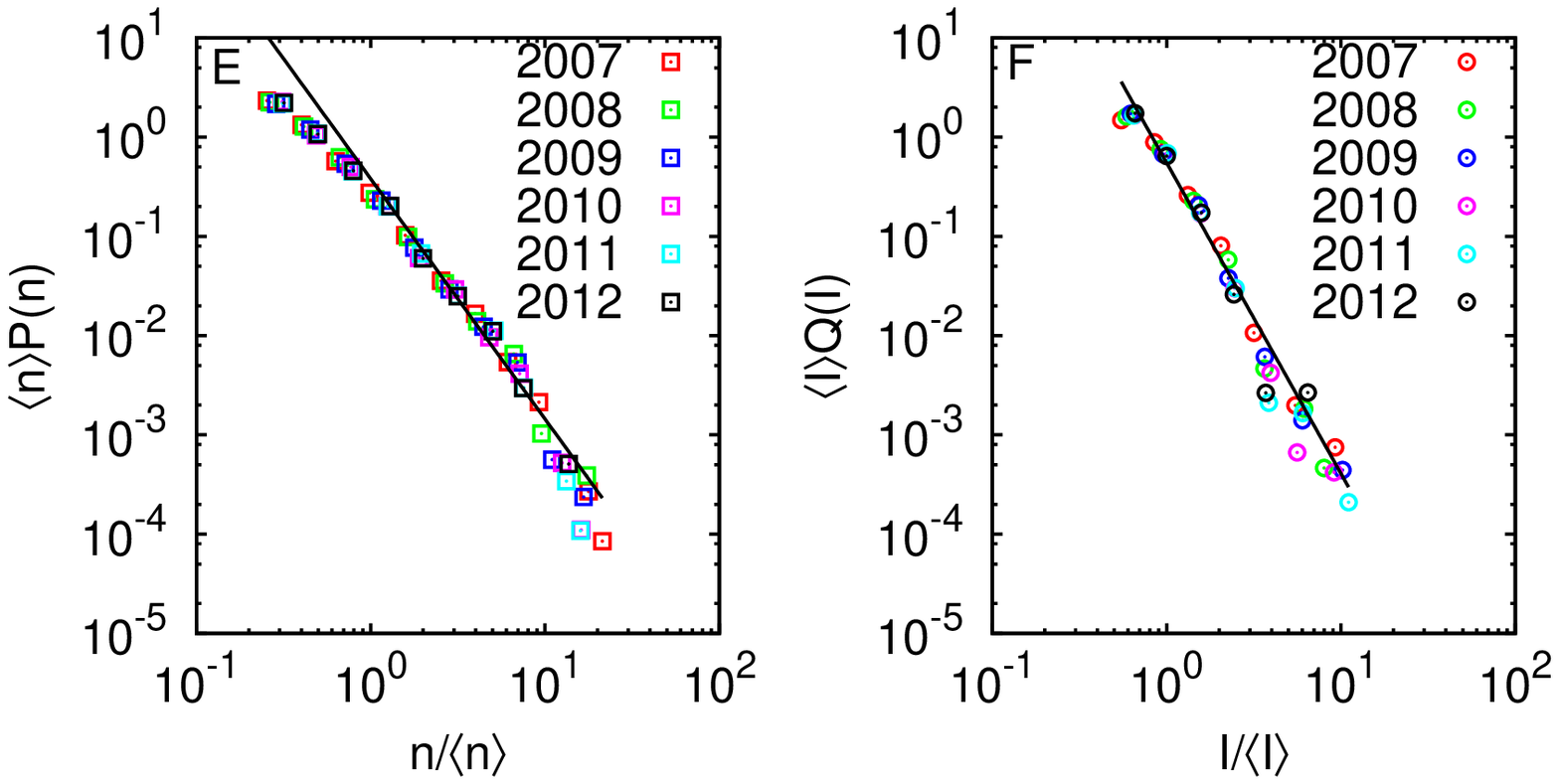}
\end{center}
     \caption{
     (A) Probability distribution of annual citations $P(n)$
   (B) its scaling collapse,  for the top $1000$ journals 
ranked according to citations.
It fits fairly well to a power law $B x^{-\gamma_n}$; 
the straight line corresponds to $\gamma_n = 2.52$ in the log-log plot; 
  (C) probability distribution of impact factor $Q(I)$ 
 (D) its scaling collapse,
 for the top $1000$ journals 
ranked according to impact factor.
The straight line corresponds to $\gamma_I = 3.16$.
The data are for SCI sets.
(E) Scaling collapse of probability distribution of annual citations $P(n)$
 for the top $1000$ journals ranked according to citations;
 power law fit with $\gamma_n = 2.42$;
 (F) Scaling collapse of probability distribution of annual citations $P(n)$
 for the top $1000$ journals ranked according to impact factor;
  power law fit with $\gamma_I = 3.50$.     The data are from SOCSCI sets.
  The figure is adapted from Ref.~\cite{khaleque2014evolution}.
}
\label{fig:dist}
 \end{figure}
Extensive studies on the historical behavior of the IF ranked 
distribution~\cite{Popescu:2003,Mansilla:2007} have established 
the behavior of the low ranked (large $k$) journals
and the precise nature of the distribution function. 
Study of  a limited sample of the top $1000$ ranked ($k \le 1000$) Science journals (SCI)
reveal that the distributions remain invariant with time, seen by rescaling the plots by their averages.
For small ranks, the citations are almost independent of the ranks implying
a cluster of journals with comparable citations that  occupy the top  positions
(Fig. \ref{fig:rank}A, B). Fitting  the curves  for $k> 10$ by $f(x) \sim x^{-b_n}$ 
it was found that $b_n = 0.70(2)$.
Similarly, for the IF, the scaled data seemed to fit to the same form with
an exponent $b_I= 0.54(1)$ (Fig. \ref{fig:rank}D, E). 
These  Zipf exponents as they are obtained from the rank plots.
For Social Science (SOCSCI), approximate power law fits are possible for the rank
plots with Zipf exponents 
$b_n= 0.70(2)$ 
and $b_I = 0.40(1)$ respectively
(Fig.~\ref{fig:rank}F, H).

The single exponent fitting on the tail does not justify the observed bending of the real data (Fig. \ref{fig:rank}B, E, H, K),
but the  curves can be as well be fitted nicely to a function with two exponents $f(x)=k(N+1-x)^{a}/x^{b}$, 
where $N$ represents the rank-order data;
$a$ and $b$ are two exponents to fit~\cite{Mansilla:2007}. For the citation data,
$b_n = 0.69(2)$ and $a_n = 0.27(2)$ (Fig. \ref{fig:rank}C). Similarly, for IF data the exponents are
$b_I = 0.56(2)$ and $a_I = 0.03(2)$ (Fig. \ref{fig:rank}F). For SOCSCI data the exponents are $b_n= 0.61(2)$, $a_n = 0.47(1)$, $b_I= 0.39(2)$, 
and $a_I = 0.06(1)$ (Fig. \ref{fig:rank}I, L). 
$a_I$ is close to zero for ({Fig. \ref{fig:rank}F, L})
implying that the power law fit is rather accurate, while for ({Fig. \ref{fig:rank}C, I}), the 
two exponents fit appears to be more appropriate.

The probability distribution of both annual citations and  impact factors showed monotonic 
decays and their tails can be fitted to power law forms.
The plots showed excellent scaling collapse for different years when 
the  probability distributions are rescaled by their averages.
For annual citations, the lower values hint towards a lognormal
but the `tail' of the distribution seems to fit to a power law~\cite{Golosovsky:2012}.
The power law exponents  are $\gamma_n$ and $\gamma_I$ 
are $2.52(1)$ and $3.16(1)$ respectively.
The Zipf exponent $b$  as obtained from a rank plot  relates to the exponent of 
the probability distribution $\gamma$  as $\gamma = 1+1/b$~\cite{Clauset:2009}.
Using the values of $b$ obtained, the values of $\gamma$   
are found to be $2.42$ and $2.85$ respectively for AC and IF, quite consistent with 
the values obtained directly from the  distributions
(Fig. \ref{fig:dist}).
The same holds true for the SOCSCI data (expected Pareto exponents are $2.42$ and 
$3.50$ from the rank plots compared to the  best fit values  $2.32(2)$ and $3.13(2)$).
However,  it is apparent that there  could be some corrections to the power law scaling.
The distribution of the annual citation rate (CR) $r$ is also broad, but non-monotonic, with a peak
around half the average value. The distributions fit well to log-Gumbel distributions~\cite{khaleque2014evolution}.

\subsection{Universality in citation distributions}
Back in 1957, Shockley~\cite{Shockley:1957} first claimed that the scientific publication rate
is dictated by lognormal distribution. Later evidences based on analysis of records for
highly cited scientists indicate that the citation distribution of individual authors follow a stretched
exponential~\cite{Laherrere:1998}. Another analysis of the data from ISI claims that the tail of the citation
distribution has a power law decay with an exponent close to $3$~\cite{Redner:1998},
What followed was  a meticulous analysis of 110 years of data from Physical Review, concluding that
most of the citation distribution fits remarkably well to a lognormal~\cite{Redner:2005}.
However, it is now well agreed that most of the distribution 
fits to a lognormal but the tail fits to a power law~\cite{peterson2010nonuniversal}.

The distribution of citations of individual publications within a single discipline is seen to be quite broad --
some papers get very little citations, a few collect huge citations, but there are many who
collect an average number of citations. The average number of citations gathered
in a discipline is strongly dependent on the discipline itself. It has been observed~\cite{Radicchi:2008} that 
if one can rescale the absolute number of citations $c$ by the average in that discipline $\langle c \rangle$,
the relative indicator $c_f=c/\langle c \rangle$ has a functional form independent
of the discipline. The rescaled probability distribution fits well to a lognormal for most of its range.

One can also ask the question that what happens for academic institutions, where the quality of
scientific output measured in terms of the total number of publications, total citations etc. can vary widely across the
world. In the popular notion, there are institutions which are known to `better' than others, and in fact, 
rankings go exist among them. One can ask if the nature of the world's best institutions' output
is different from the more average ones. It is found that all institutions have a wide distribution
of citations to their publications, with a functional form that is independent of their rankings~\cite{chatterjee2014universality}.
To see this, one has to use a relative indicator $c_f=c/\langle c \rangle$ again. The scaling function
fits to a lognormal for most of its range, while the highest cited papers deviate to fit to a power law tail.

Now, how does the same look at the level of journals? It is quite well known in popular perception
that journals present a wide variety in terms of impact factor and annual 
citations, as also reported in critical studies~\cite{khaleque2014evolution,Popescu:2003,Mansilla:2007}.
The natural question to ask is whether the universality that exists across disciplines or academic institutions
prevail also  for journals. It seems that this is not the case in reality.
In fact, it is seen that there are at least two classes of journals, a \textit{General} class,
which comprises the most popular standard journals, and an \textit{Elite} class
consisting a very small group of highly reputed journals characterized by high impact factor 
and average citations~\cite{chatterjee2014universality}. The former class is characterized by a lognormal distribution in the bulk,
which the latter is characterized by a strong tendency of divergence as $1/c$ at small values of citations, with no clear
indication of lognormal bulk. However, for both these classes, the distribution of the
highest cited papers decay a power law tail with an exponent close to $3$.

\section{Networks}
\label{sec:networks}
Networks have been a subject of intense study in the last 2 decades, and drawn researchers 
from a variety of disciplines to study the structural, functional, dynamical and various
aspects of them, uncovering new patterns and understanding interesting phenomena
from physics, biology, computer science and social sciences~\cite{newman2006structure}.
The studies of the internet and the web has led to engineering faster
routing strategies and developing efficient search engines, biological networks 
have given insight into the functional elements in bio-chemical processes inside
organisms, provided tools for complex experiments, that in social networks
has bettered our understanding about spreading of innovations, rumors
and even epidemics, leading to devising strategies to make them either more efficient 
or less efficient according to the case it may be.

Studies of the World Wide Web~\cite{albert1999internet,broder2000graph}, the internet~\cite{faloutsos1999power},
citation networks~\cite{Price:1965,Redner:1998}, 
email network~\cite{ebel2002scale}, network of human sexual contacts~\cite{liljeros2001web}
all show power law tail in the degree distributions for a large range of values, similar to 
metabolic networks~\cite{jeong2000large} and protein-protein interaction network~\cite{uetz2000comprehensive}.

Studies on massive and popular social networks like Facebook~\cite{ugander2011anatomy}
has shown broad distributions with tails resembling power laws.
Experiments done by setting up social networks of communication have shown that the network slowly
evolves and the shape of the degree distribution stabilizes with time~\cite{kossinets2006empirical},
with characteristics similar to other massive networks, such as clustering and broad degree distribution
with power law tail. Since then, almost all possible online social networks have been analyzed,
and most of them have been found to have a power law degree distribution (see e.g., Ref.~\cite{li2014sparse}).

\section{Measuring inequality}
\label{sec:measure}
Socio-economic inequalities can be quantified in several ways. The most common measures are absolute,
in terms of indices, e.g., Gini~\cite{gini1921measurement}, Theil~\cite{theil1967economics} indices.
Alternatively one can use  a relative measure,
by considering the probability distributions of various quantities, whereas the  indices
can be computed easily from the distributions themselves. The quantity in question in the socio-economic context
usually displays a broad distribution, like lognormals, power-laws or their combinations.
For instance, the distribution of income is usually an exponential followed by a power 
law~\cite{druagulescu2001exponential} (see Ref.\cite{chakrabarti2013econophysics} for other examples).

\begin{figure}[t]
\begin{center}
\includegraphics[width=9.0cm]{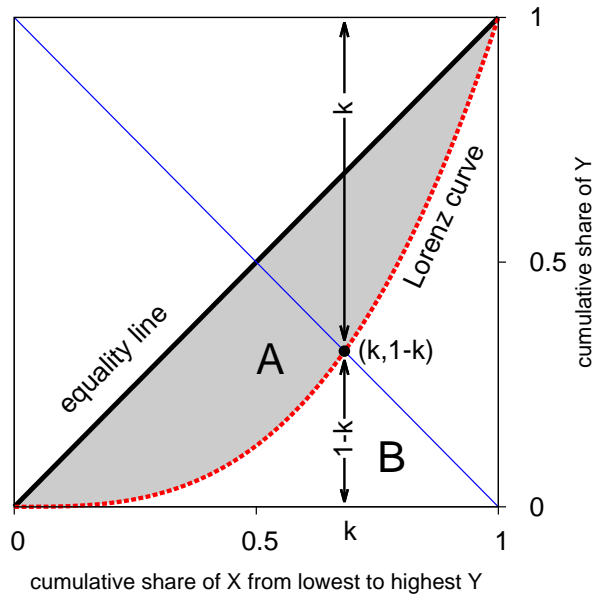}
\end{center}
\caption{
Schematic representation of 
Lorenz curve, Gini index $g$ and $k$ index.
The dashed red line stands for the Lorenz curve and the black solid line represents perfect equality.
The area enclosed by the equality line and the Lorenz curve is $A$ and that below the Lorenz curve is $B$.
The Gini index is given by $g=A/(A+B)$.
The $k$ index is given by the abscissa of the intersection point of the Lorenz curve and $Y=1-X$.
}
\label{fig:fg_pic_Lorenz}
\end{figure}

The Lorenz curve~\cite{Lorenz} is a function representing the cumulative proportion $X$ of 
(ordered from lowest to highest) individuals 
in terms of the cumulative proportion of their sizes $Y$.
$X$ can be anything like income or wealth, citation, votes, city population etc. 
The Gini index ($g$) is commonly defined as the ratio of (i) the area enclosed between
the equality line and the Lorenz curve, and (ii) the area below the equality line.
If the area between 
(i) the Lorenz curve and the equality line is given by $A$, and that  
(ii) under the Lorenz curve is given by $B$  (See Fig.~\ref{fig:fg_pic_Lorenz}),
the Gini index is simply defined as $g=A/(A+B)$.
It is a very common  measure to quantify socio-economic inequalities. 
Ghosh et al.~\cite{ghosh2014inequality}
recently introduced a new  measure,  
called `$k$ index' (where `$k$' stands for the extreme socio-economic inequalities in Kolkata),
defined as the fraction $k$ such that  $(1-k)$ fraction 
of people or papers possess $k$ fraction of income/wealth or citations respectively.

When the probability distribution is described by an appropriate parametric function, one can derive, using mathematical techniques,
these inequality measures as a function of those parameters analytically. 
Several empirical evidence show that the distributions can be put into a few types, most of which turn out to be a
mixture of two distinct parametric functions with a single crossover point~\cite{inoue2014measuring}
\begin{equation}
P(m) = F_{1}(m)\theta(m,m_{\times})+
F_{2}(m)\Theta (m-m_{\times}),
\end{equation}
where $\theta (m,m_{\times}) \equiv \Theta (m)-\Theta (m-m_{\times})$,
$m_{\times}$ is the crossover point and  $F_1$, $F_2$ being the two functions.
For these, it is possible to compute the general form of the Lorenz curve, Gini index $g$ and $k$ index.
It is also easy to check the values obtained from empirical calculations with those from analytical expressions for consistency. 
By minimizing the gap between the empirical and analytical values of
the inequality measures, one can numerically enumerate of the crossover point which is usually determined by eye estimates.


\section{How to deal with inequality?}
\label{sec:deal}
\begin{quotation}
 \textit{So distribution should undo excess, and each man have enough.}
\begin{flushright}
 [King Lear, Act 4, Scene 1] -- William Shakespeare, English playright and writer (1564-1616)
\end{flushright}
 \end{quotation}
In this section, we will not discuss socio-political theories such as \textit{socialism}
in it different forms, and rather give an objective view of generalized
processes that can be adopted to decrease inequalities.
 
One of the long studied problems that deals with the issue of 
socio-economic inequality is that of efficient allocation of resources.
An inefficient allocation process may lead to unequal distribution of
resources which in course of time goes through a reshuffling and multiplicative
process, and finally result in a very skew distribution.
We will discuss in the following, how statistical physics
can help in modeling processes that produce fairly equal distributions.

Most resources are limited in supply, and hence, efficiently allocating them is
therefore of great practical importance in many fields. Resources may
be either physical (oil, CPUs, chocolates) or immaterial (time,
energy, bandwidth), and allocation may happen instantaneously or over a
long period. Thus, there is no universal method that solves all
allocation problems. In addition, the variety of situations that
people or machines (generically called ``agents'') face, require
specific modeling in a first step. However, two features connect all
these situations: when resources are scarce, the agents compete for
them; when competing repeatedly, the agents learn and become
adaptive. Competition in turn has two notable consequences. 
The agents have a strong incentive to think and act as differently as 
possible~\cite{Arthur}, 
that is, to become heterogeneous. Additionally,
competition implies interaction, because the share of resources that
an agent obtains depends on the actions of other agents.

While multi-agent modeling in the context of Game theory helps,
even simpler modeling using the framework of statistical 
physics~\cite{chakraborti2013statistical}
helps in a deeper understanding of the problem using a framework which
is very different from those traditionally used by mainstream social scientists.
While some models have already been proposed and efficiently utilized in certain fields,
there is further  need of  developing novel and efficient models for a variety of problems.

The optimal allocation of resources is an issue of utmost concern in economic systems. 
Formalized as the simultaneous maximization of the utility of each member of the 
economy over the set of achievable allocations, the main issue is that individuals have typically conflicting
goals, as the profit of one goes against that of the others. This makes the nature of the problem 
conceptually different from optimization problems where usually a single objective function has to be maximized. 
Markets, under some conditions, can solve this problem efficiently -- prices adjust in a self-organized manner
to reflect the true value of the good.

There have been recent attempts to model and describe resource allocation 
problems in a collection of individuals within a statistical mechanics approach.
The focus is on competitive resource allocation models where the   decision process is
fully decentralized, that is, communication between the agents is not explicit.
Interaction plays a crucial role and  give rise to collective
phenomena like broad distribution of
fluctuations, long
memory and even phase transitions. Additionally, the agents
usually act very differently from one another, and sometimes have the ability
to change actions according to the need their goals. This implies strong
heterogeneity and non-equilibrium dynamics, ingredients which 
are extremely appealing to physicists, who possess the tools and concepts that are 
able to analyze and possibly solve the dynamics of such systems.
The essential difference lies in the behavior of the constituent units,
while particles, electrons cannot think and act, social units sometime do,
which adds to the complexity.

Let us consider a population of $N$  agents and $R$ resources, which they try to exploit.
Generically, $R$ denotes the number of possible choices of the agents, which naturally means $R\ge2$. 
For the simplest case $R=2$, agents must choose which resource to exploit.
The El Farol Bar Problem (EFBP)~\cite{Arthur}: $N$ customers compete for $L<N$ seats at the bar. At every time step, they must choose whether to go to the bar or to stay at home. The Minority Game (MG) \cite{CZ97} simplifies the EFBP in many respects by taking $L=N/2$. 
The Kolkata Paise Restaurant problem (KPR) assumes that the number of resources scales with $N$, in which $R$ restaurants
have a capacity to serve only one customer each, so that the agents try to remain alone as often as possible~\cite{Chakrabarti2009}. 

The KPR~\cite{Chakrabarti2009,Ghosh2010} problem is similar to various adaptive games 
(see \cite{Challet2004}) but uses one of the simplest scenario to model resource utilization
in a toy model paradigm. 
The simplest version has
$N$  agents (customers) simultaneously choosing equal number  $R$ ($=N$) of restaurants,
each of which serve only one meal every evening, and hence, showing up in a place with more people 
gives less chance of getting food.
\textit{Utilization} is measured by the fraction
of agents $f$ getting a meal or equivalently, by measuring the complimentary quantity: 
the fraction ($1-f$) of meals wasted, because some of the restaurants do not get any customer at all.
An entirely random occupancy algorithm sets a benchmark of $f= 1-1/e \approx 0.63$. However, 
a crowd-avoiding algorithm~\cite{Ghosh2010} can improve the utilization to around $0.8$.
If ones varies the ratio of  agents to restaurants $(N/R)$ below unity, 
one can observe a phase transition -- from  an `active phase' characterized
by a finite fraction of restaurants  $\rho_a$  with more than one agent, and an `absorbed phase',
where  this quantity $\rho_a$ vanishes~\cite{Ghosh2012}. Adapting the  same crowd avoiding strategy
in a version of the Minority Game where the extra information
about the crowd was provided, one could convergence to the steady state achieve in a very small time, 
$O( \log \log N)$~\cite{Dhar2011}. In another modification to this problem~\cite{Biswas2012},
one could observe a phase transition depending on the amount of information that is shared.
The main idea for the above studies was to use iterative learning to find simple algorithms that could
achieve a state of maximum utilization in a very small time scale.
Although this review touches upon topics seemingly discrete from one another, some of the processes are 
interlinked, e.g., resource utilization can be seen as a key ingredient to city growth
and the broad distribution of city sizes~\cite{ghosh2014zipf}.

\section{Discussions}
\label{sec:discuss}
Socio-economic inequalities have been around, manifested in several forms, since history.
With time, the nature and extent of these have changed, sometimes for good, but mostly for worse.
This is traditionally a subject of study of social sciences, and scholars have been looking upon the 
causes and effects from a sociological perspective, and trying to understand the consequences
on the economics. In reality this is not so simple,  at times the latter is what is responsible for the 
effect on the former, making the cause-effect interpretation much more complex.

At the opposite end, had the world been very equal, it would have been difficult to 
compare the extremes, separate the good from the bad, hardly any leaders people would look up to,
lack stable governments if there were almost equal number of political rivals, etc.

Recently, there has been a lot of concern about the increase of inequality in income and wealth,
as seen from different measurements~\cite{piketty2014capital}, and has in such, renewed the interest on this subject
among the leading social scientists across the globe. However, energy use has been
observed to become much more equal with time~\cite{lawrence2013global}.
In the economics front, how inequalities affect financial markets,
firms and their dynamics, and vice versa is an important area to look into.
Objects which are directly affected by economy, industrialization and rapidly growing technologies, 
in terms of social organization of individual entities, as in urban systems, are becoming 
important areas of study.
In climbing up and down the social ladder~\cite{mervis2014tracking,bardoscia2013social}
is something difficult to track, until recent surveys which provide
some insight into the dynamics.
Many deeper and important issues in society~\cite{neckerman2004social} 
still needs attention in terms of inequality research,
available data and its further analysis can bring out hidden patterns which may be used 
to encounter those situations. The main handicap is the lack of data of good quality,
and the abstractness of the issues, which may not always be easily amenable to
statistical modeling.

Physicists' interests are mostly concentrated on subjects which are 
amenable to macroscopic or microscopic modeling, where tools of statistical physics
prove useful in explaining the emergence of broad distributions. A huge body of
literature has been already developed, containing serious attempts
to understand socio-economic phenomena, branded under \textit{Econophysics}
and \textit{Sociophysics}~\cite{chakrabarti2007econophysics}.
The physics perspective brings new ideas and an alternative outlook
to the traditional approach taken by social scientists,
and is seen in the increasing collaborations across disciplines~\cite{lazer09}.

\begin{acknowledgments}
 This mini review is a starting point of a larger material to be written up. 
 Discussions with V.M. Yakovenko were extremely useful in planning the contents.
 The author thanks S. Biswas, A.S. Chakrabarti and B.K. Chakrabarti for comments,
 and acknowledges collaborations with 
 F. Becattini, S. Biswas,  A.S. Chakrabarti, B.K. Chakrabarti,, A. Chakraborti, S.R. Chakravarty,  D. Challet, A.K. Chandra,
  D. DeMartino,  S. Fortunato, A. Ghosh,
  J-I. Inoue, A. Khaleque, S.S. Manna, M. Marsili, M. Mitra, M. Mitrovi\'{c}, 
    T. Naskar, R.K. Pan, P.D.B. Parolo, P. Sen,
 S. Sinha, Y-C. Zhang on various projects.
\end{acknowledgments}

\bibliographystyle{unsrt}
\bibliography{ref.bib}

\end{document}